\documentclass[aps,pra,reprint,groupedaddress,longbibliography]{revtex4-1}

\pdfoutput=1

\usepackage{graphicx,color}
\usepackage{amsmath, amssymb}
\usepackage{longtable}
\usepackage{natbib}

\usepackage{multirow}

\begin{document}
\title{
Orbital disproportionation of electronic density - a universal feature of alkali-doped fullerides 
}
\author{Naoya Iwahara}
\author{Liviu F. Chibotaru}
\affiliation{Theory of Nanomaterials Group, 
Katholieke Universiteit Leuven, 
Celestijnenlaan 200F, B-3001 Leuven, Belgium}
\date{\today}

\begin{abstract}
Alkali-doped fullerides A$_n$C$_{60}$ show a remarkably wide range of electronic phases in function of 
A = Li, Na, K, Rb, Cs 
and the degree of doping, $n=$ 1-5. 
While the presence of strong electron correlations is well established, recent investigations give also evidence for 
dynamical Jahn-Teller instability in the insulating and the metallic phase of A$_3$C$_{60}$. 
To reveal the interplay of these interactions in fullerides with even $n$, we address the electronic phase of 
A$_4$C$_{60}$ with accurate many-body calculations within a realistic electronic model including all basic interactions 
extracted from first principles. We find that the Jahn-Teller instability is always realized in these materials too. 
More remarkably, in sharp contrast to strongly correlated A$_3$C$_{60}$, 
A$_4$C$_{60}$ displays uncorrelated band-insulating state despite pretty similar interactions present in both fullerides. 
Our results show that the Jahn-Teller instability and the accompanying orbital disproportionation of electronic density in 
the degenerate LUMO band is a universal feature of fullerides.
\end{abstract}

\maketitle

\section{Introduction}
\label{Sec:Introduction}
The understanding of electronic phases of alkali-doped fullerides A$_n$C$_{60}$ is a long standing and challenging task for material scientists \cite{Gunnarsson2004}. 
The prominent feature of these narrow-band molecular materials is the coexistence of strong intrasite Jahn-Teller (JT) effect 
with strong electron correlation, which underlies the unconventional superconductivity in A$_3$C$_{60}$ 
\cite{Gunnarsson1997RMP, Ganin2008, Takabayashi2009, Capone2009, Ganin2010, Ihara2010, Ihara2011, Nomura2016}
and a broad variations of electronic properties in this series of materials in function of the size of alkali ions, 
and the degree of their doping \cite{Tanigaki1991, Tanigaki1992, Winter1992, Murphy1992, Kiefl1992}. 
External pressure and insertion of neutral spacers add new possibilities for the engineering of their electronic phases 
\cite{Rosseinsky1993, Durand2003, Ganin2006}. 
This was recently demonstrated for the Cs$_3$C$_{60}$ fulleride, which undergoes transitions from Mott-Hubbard (MH) antiferromagnet to a high temperature superconductor ($T_c =$38 K) and then to strongly correlated metal under external pressure
\cite{Ganin2008, Takabayashi2009, Ganin2010, Ihara2010, Ihara2011}. 

Signs of JT effect in alkali-doped fullerides were inferred from 
NMR \cite{Brouet2001, Potocnik2014}, IR \cite{Klupp2006, Klupp2012}, and EELS \cite{Knupfer1996, Knupfer1997} spectra,
and STM \cite{Wachowiak2005, Dunn2015} in various compounds.
Recently, the parameters governing the complex JT interaction on fullerene anions have been firmly established
\cite{Iwahara2010, LaflammeJanssen2010, Faber2011}, 
which opened the way for accurate theoretical investigation of the electronic states in fullerides. 
It was found that in the MH insulating phase of cubic fullerides such as Cs$_3$C$_{60}$ at ambient pressure, 
the para dynamical JT effect is realized as independent pseudorotations
of JT deformations at each C$_{60}$ site \cite{Iwahara2013}. 
The same para dynamical JT effect was found in the metallic phase of A$_3$C$_{60}$ close to MH transition, while the 
pseudorotation of JT deformation at different sites are expected to be correlated with further departure from the 
MH transition due to the increase of the band energy \cite{Iwahara2015}. 
These findings have found confirmation in a very recent investigation of Cs$_3$C$_{60}$ fulleride, 
showing an almost unchanged IR spectrum on both sides in the vicinity of MH metal-insulator transition,
while displaying its significant variation when the material was brought deeper into the metallic phase \cite{Zadik2015}. 
Moreover, our calculations have also 
shown that the metallic phase in these systems exhibits an orbital disproportionation of electronic density 
as a result of the dynamical JT instability \cite{Iwahara2015}.  

This successful theoretical approach is applied here for the investigation of the electronic phase in the A$_4$C$_{60}$ fullerides, 
containing an even number of doped electrons per site. 
We find that these materials exhibit a dynamical JT instability too. 
As in A$_3$C$_{60}$, the ground state of A$_4$C$_{60}$ displays again the orbital disproportionation of electronic density, thus identifying it as a universal key feature of the electronic phases of alkali-doped fullerides.

\section{Diagram of Jahn-Teller instability in A$_4$C$_{60}$}
\label{Sec:Energy}
It is well established that the $t_{1u}$ lowest unoccupied molecular orbital (LUMO) 
band mainly defines the electronic properties of fullerides \cite{Gunnarsson2004}. 
Following the recent treatment of A$_3$C$_{60}$ \cite{Iwahara2015}, 
we consider all essential interaction in this band including the one-electron, the bielectronic and the vibronic contributions:
\begin{eqnarray*}
\hat{H} &=& \hat{H}_{\text{t}} + \hat{H}_{\text{bi}} + \hat{H}_{\text{JT}}, 
\nonumber\\
\hat{H}_{\rm t} &=& \sum_{\mathbf{m},\Delta \mathbf{m}}\sum_{\lambda \lambda' \sigma} 
                     t_{\lambda \lambda'}^{\Delta \mathbf{m}} 
                     \hat{c}_{\mathbf{m}+\Delta \mathbf{m} \lambda\sigma}^\dagger \hat{c}_{\mathbf{m} \lambda'\sigma}, 
\nonumber\\
 \hat{H}_{\rm bi} &=& \frac{1}{2}\sum_\mathbf{m}\sum_{\lambda \sigma} \left[ 
                      U_\parallel \hat{n}_{\mathbf{m} \lambda \sigma} \hat{n}_{\mathbf{m} \lambda -\sigma}  
                      \right.
\nonumber\\
                  &+& 
                      U_\perp \sum_{\lambda' (\ne \lambda)\sigma'} 
                      \hat{n}_{\mathbf{m} \lambda \sigma} \hat{n}_{\mathbf{m} \lambda' \sigma'}  
                   - J \sum_{\lambda' (\ne \lambda)} \left(
                      \hat{n}_{\mathbf{m} \lambda \sigma} \hat{n}_{\mathbf{m} \lambda' \sigma}  
                      \right.
\nonumber\\
                  &-& \hat{c}_{\mathbf{m} \lambda \sigma}^\dagger \hat{c}_{\mathbf{m} \lambda' \sigma}
                      \hat{c}_{\mathbf{m} \lambda -\sigma}^\dagger \hat{c}_{\mathbf{m} \lambda' -\sigma}  
\nonumber\\
                  &-&
                      \left.
                      \left.
                      \hat{c}_{\mathbf{m} \lambda \sigma}^\dagger \hat{c}_{\mathbf{m} \lambda' \sigma}
                      \hat{c}_{\mathbf{m} \lambda' -\sigma}^\dagger \hat{c}_{\mathbf{m} \lambda -\sigma}  
                 \right)
                 \right],
\end{eqnarray*}
\begin{eqnarray}
 \hat{H}_{\rm JT} &=& \sum_{\mathbf{m}} 
                    \hslash \omega
                    \left[
                     \sum_\gamma \frac{1}{2} 
                     \left(
                     p_{\mathbf{m}\gamma}^2 
                     +
                     q_{\mathbf{m}\gamma}^2
                     \right)
                    \right.
\nonumber\\
                    &+&
                    \left.
                     g \sum_{\lambda \lambda'\sigma} \sum_\gamma 
                    G_{\lambda \lambda'}^\gamma
                    \hat{c}_{\mathbf{m}\lambda \sigma}^\dagger 
                    \hat{c}_{\mathbf{m}\lambda' \sigma}
                    q_{\mathbf{m}\gamma} 
                   \right],
\label{Eq_1}
\end{eqnarray}
where, ${\mathbf{m}}$ denote the fullerene sites, 
$\Delta \mathbf{m}$ the neighbours of site $\mathbf{m}$, 
$\lambda, \lambda'$ the $t_{1u}$ LUMO orbitals ($x,y,z$) on each C$_{60}$,
$\sigma,\sigma'$ the spin projections,
$\hat{c}_{\mathbf{m} \lambda \sigma}$ and $\hat{c}_{\mathbf{m} \lambda \sigma}^\dagger$ are 
annihilation and creation operators of electron, respectively, 
$\hat{n}_{\mathbf{m} \lambda \sigma} = \hat{c}^\dagger_{\mathbf{m} \lambda \sigma} \hat{c}_{\mathbf{m} \lambda \sigma}$,
$q_{\mathbf{m}\gamma}$ and $p_{\mathbf{m}\gamma}$ are the normal vibrational coordinate for the 
$\gamma$ component of the $h_g$ mode ($\gamma=\theta, \epsilon, \xi, \eta, \zeta$) and its conjugate momentum, respectively,
and $G_{\lambda \lambda'}^\gamma$ is Clebsch-Gordan coefficient. 
The transfer parameters $t_{\lambda \lambda'}^{\Delta \mathbf{m}}$ 
of $\hat{H}_{\text{t}}$ have been extracted from density functional theory (DFT) calculations 
(see Ref. \cite{Iwahara2015} for K$_3$C$_{60}$, Supplementary Materials and 
Fig. \ref{Fig5}(A) for K$_4$C$_{60}$). 
The frequency $\omega$ and the orbital vibronic coupling constant $g$ 
for an effective single-mode JT model of C$_{60}^{n-}$ have been calculated in Ref. \cite{Iwahara2013}.
The phonon dispersion was neglected because it is weak in fullerides \cite{Gunnarsson2004}. 
The projection of the bielectronic interaction in the $t_{1u}$ LUMO band onto intrasite 
Hamiltonian ($\hat{H}_{\text{bi}}$) 
is an adequate approximation due to strong molecular character of fullerides \cite{Gunnarsson2004}. 
The intrasite repulsion parameters $U_\parallel$ and $U_{\perp}$, obeying the relation $U_\parallel -U_{\perp} =2J$, 
are strongly screened: first, by high-energy interband
electron excitations reducing their value from 3 eV to ca 1 eV \cite{Nomura2012} and, 
second, by intra $t_{1u}$-band excitations. 
The latter can further reduce $U_\parallel$ and $U_{\perp}$ several times \cite{Nomura2012}, however, 
the extent of this screening strongly depends on the character of the correlated $t_{1u}$ band and can, 
therefore, be assessed only in a self-consistent fashion. 
On the other hand, the vibronic coupling to the $h_g$ modes, representing a quadrupolar perturbation, is hardly screened.
The same for the Hund's rule coupling $J$, for which we take the calculated molecular value \cite{Iwahara2013}. 
We leave $U_\parallel$ as the only free parameter of the theory. 
\begin{figure}[bt]
\begin{center}
\begin{tabular}{l}
(A)\\
\includegraphics[width=7cm, bb=0 0 3965 2576]{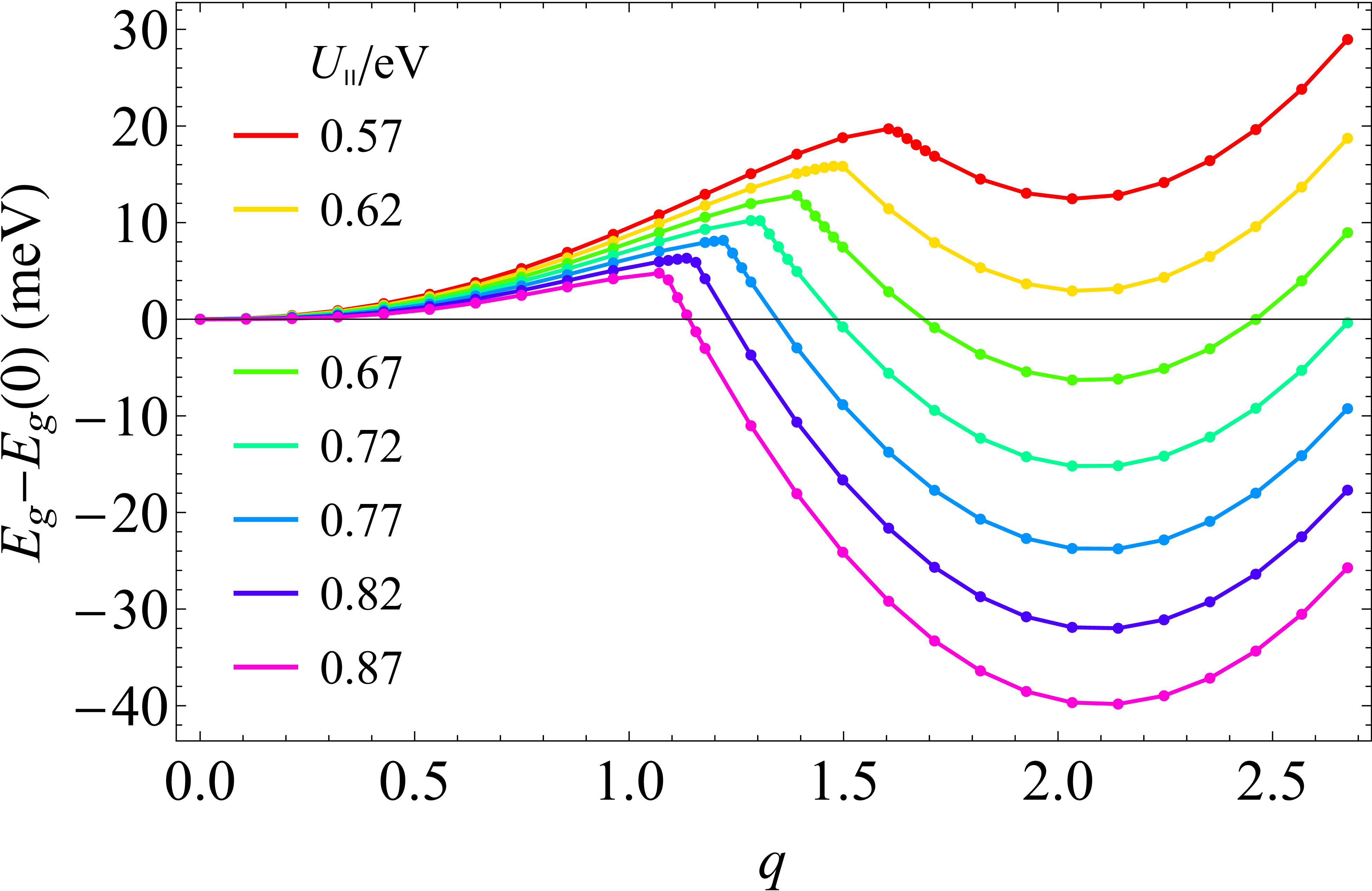}
\\
(B)
\\
\includegraphics[width=7cm, bb=0 0 3417 2861]{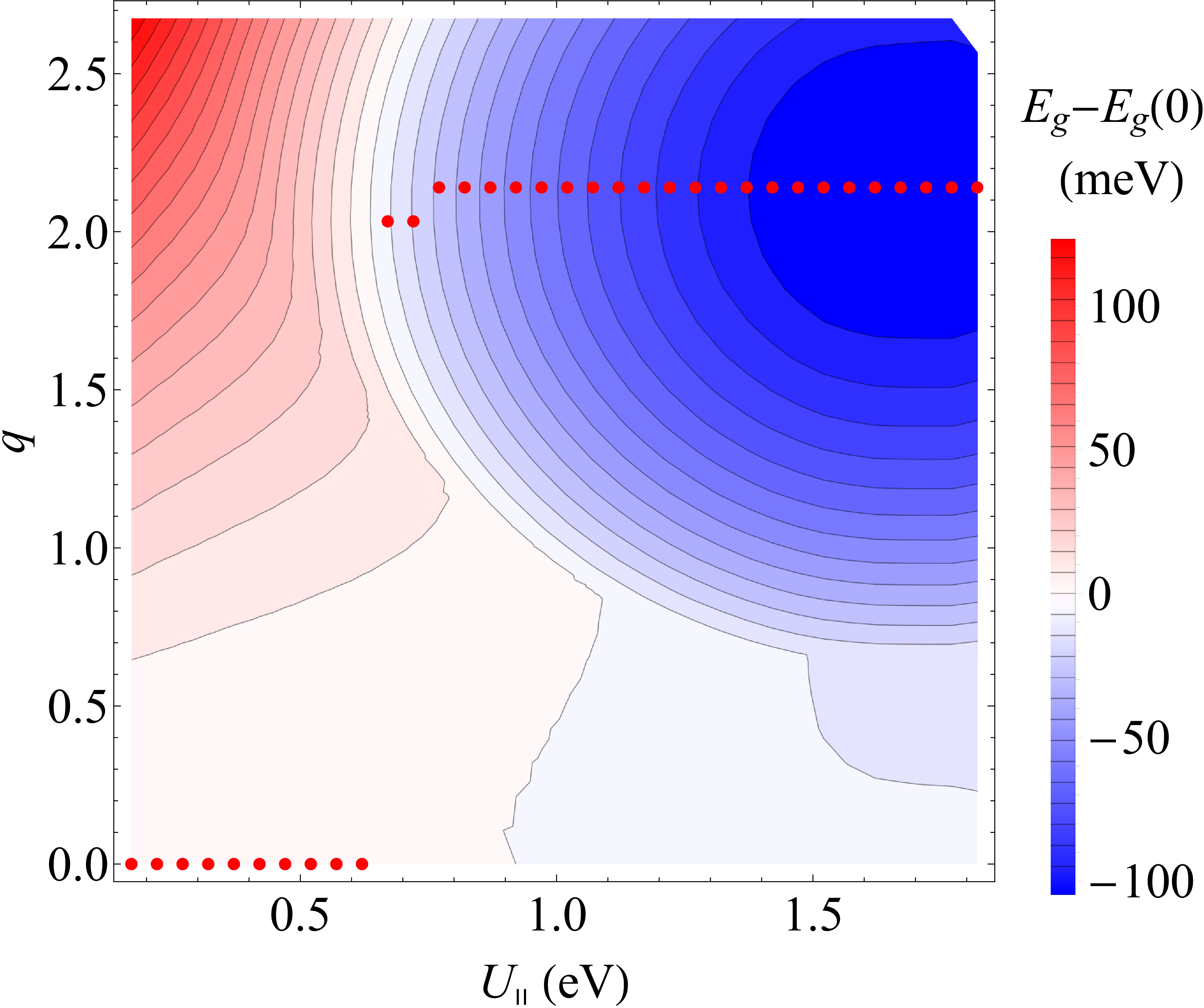}
\end{tabular}
\end{center}
\caption{
(A) Total energy $E_g (q)$ of the ground electronic phase of a A$_4$C$_{60}$ with cubic band dispersion (see the text) 
as function of amplitude of static JT distortion for several values of $U_\parallel$. 
(B) A two-dimensional plot of $E_g (q,U_\parallel)$. 
Red and blue regions stand for positive and negative values proportional to the intensity of the color. 
The red points show the amplitude of spontaneous static JT distortion in function of $U_\parallel$.
In both figures, the energy at $q=0$, $E_g(0)$, is subtracted from $E_g(q)$ for each $U_\parallel$.
}
\label{Fig1}
\end{figure}

The ground state has been calculated within a self-consistent Gutzwiller approach,
which proved to be successful for the investigation of A$_3$C$_{60}$ \cite{Iwahara2015}. 
To unravel the role played by JT interactions in the ground electronic phase in A$_4$C$_{60}$,
we first consider the case of a face centered cubic (fcc) $\hat{H}_{\text{t}}$ as in A$_3$C$_{60}$, 
the corresponding bands being populated by four electrons per site. 
Figure \ref{Fig1}(A) shows the calculated total energy as function of the amplitude $q$ 
of static JT distortions of $h_g\theta$ type on fullerene sites \cite{Auerbach1994, OBrien1996}. 
As in the case of A$_3$C$_{60}$ \cite{Iwahara2015},
the energy curve $E_g(q)$ has two minima, one at the undistorted configuration $q=0$ 
and the other at a value $q_0$ approximately corresponding to the equilibrium distortion in an isolated C$_{60}^{4-}$ 
(see the Supplementary Materials). 
For $U_{\parallel}$ smaller than the critical value $U_{c} \approx$ 0.64 eV, the static JT distortion is quenched, $q=0$. 
At $U_{\parallel} > U_{c}$ the JT distortion reaches its equilibrium value, $q_0$.
The full diagram of the total energy $E_g (q,U_\parallel)$ is shown in Fig. \ref{Fig1}(B). 

\begin{figure*}[bt]
\begin{center}
\begin{tabular}{lll}
(A) & ~~ & (B) \\
\includegraphics[width=7cm, bb= 0 0 3931 2618]{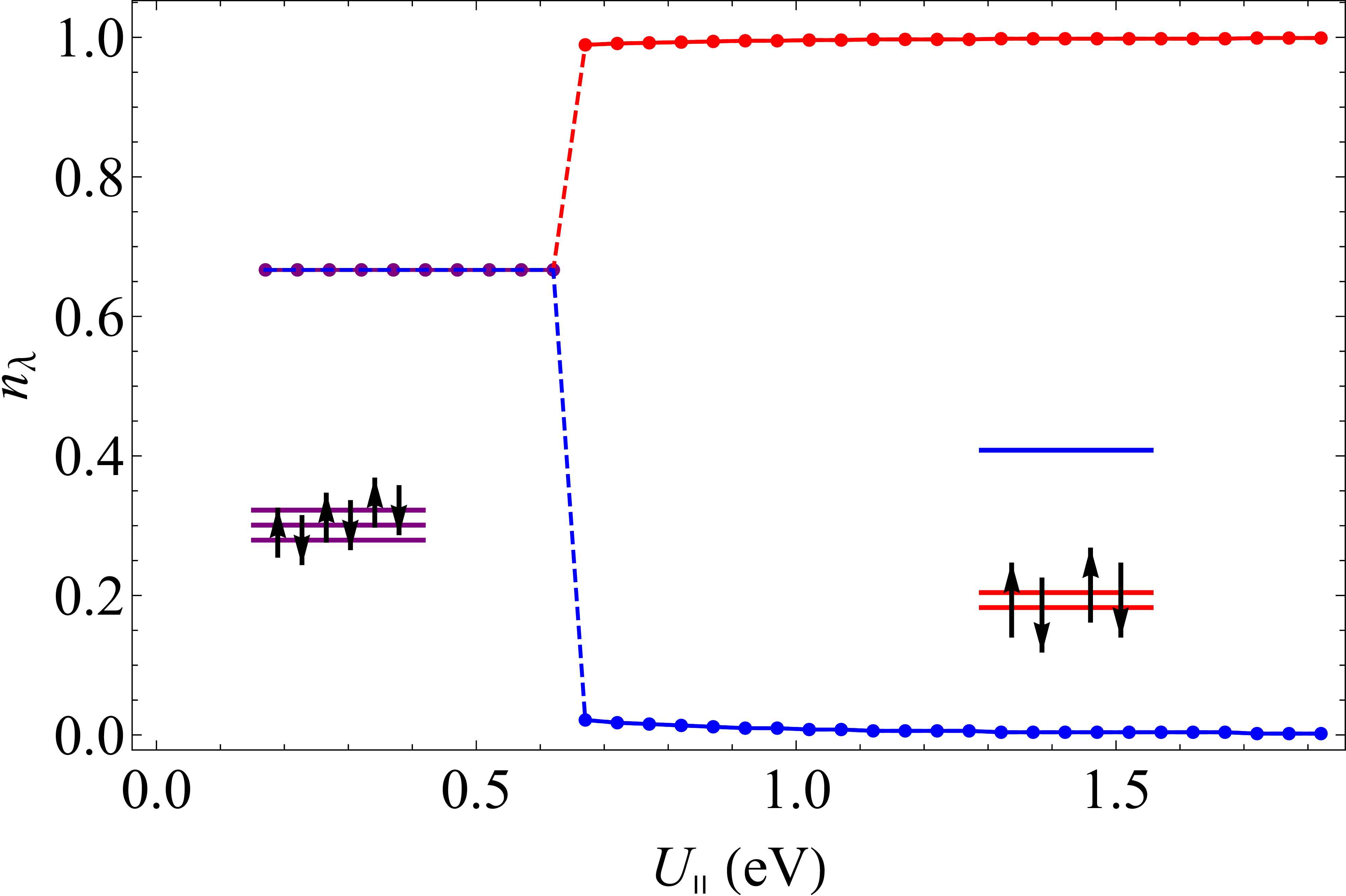}
&&
\includegraphics[width=7cm, bb= 0 0 3931 2618]{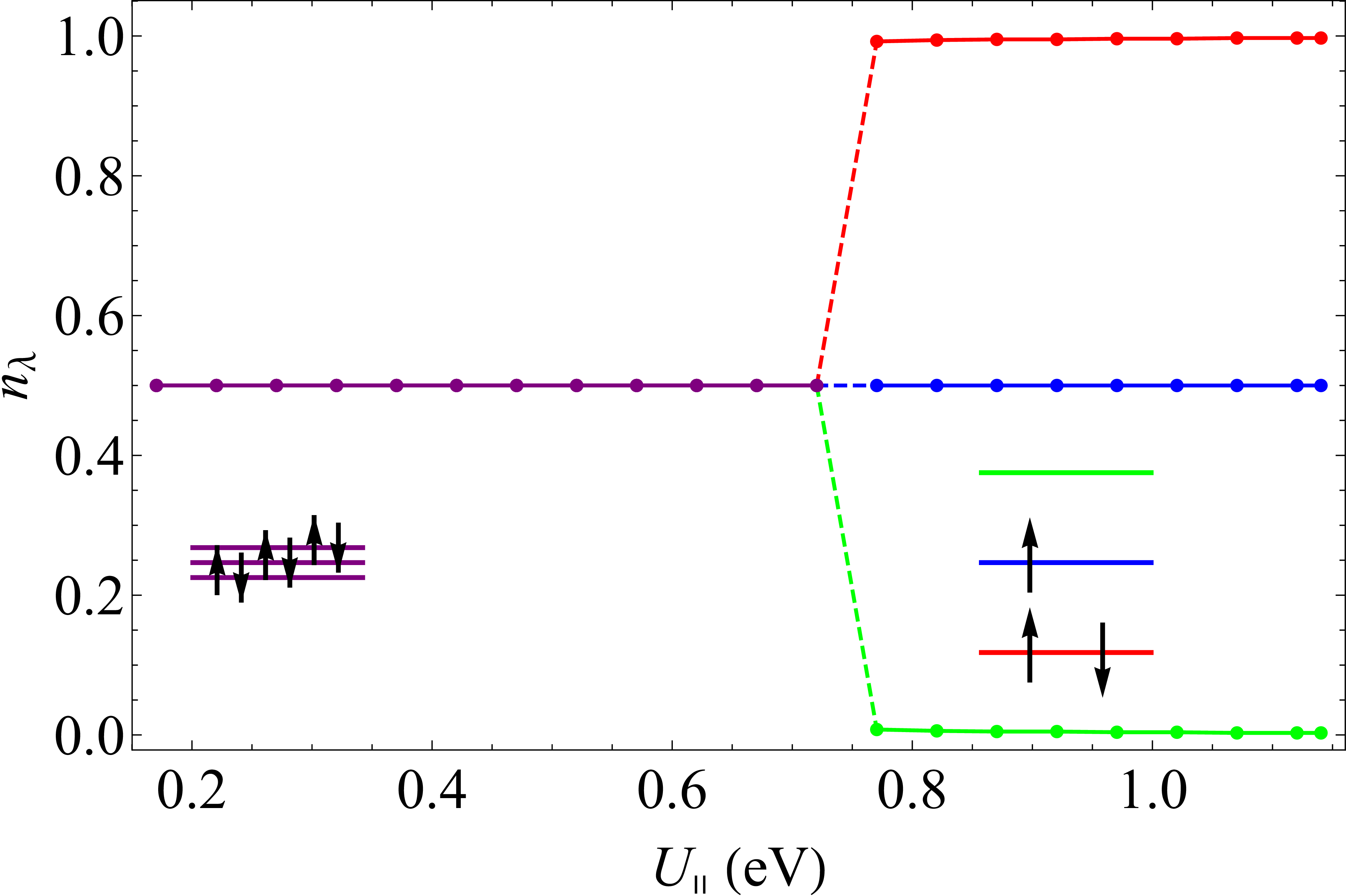}
\\
(C) && (D) \\
\includegraphics[width=7cm, bb= 0 0 3931 2569]{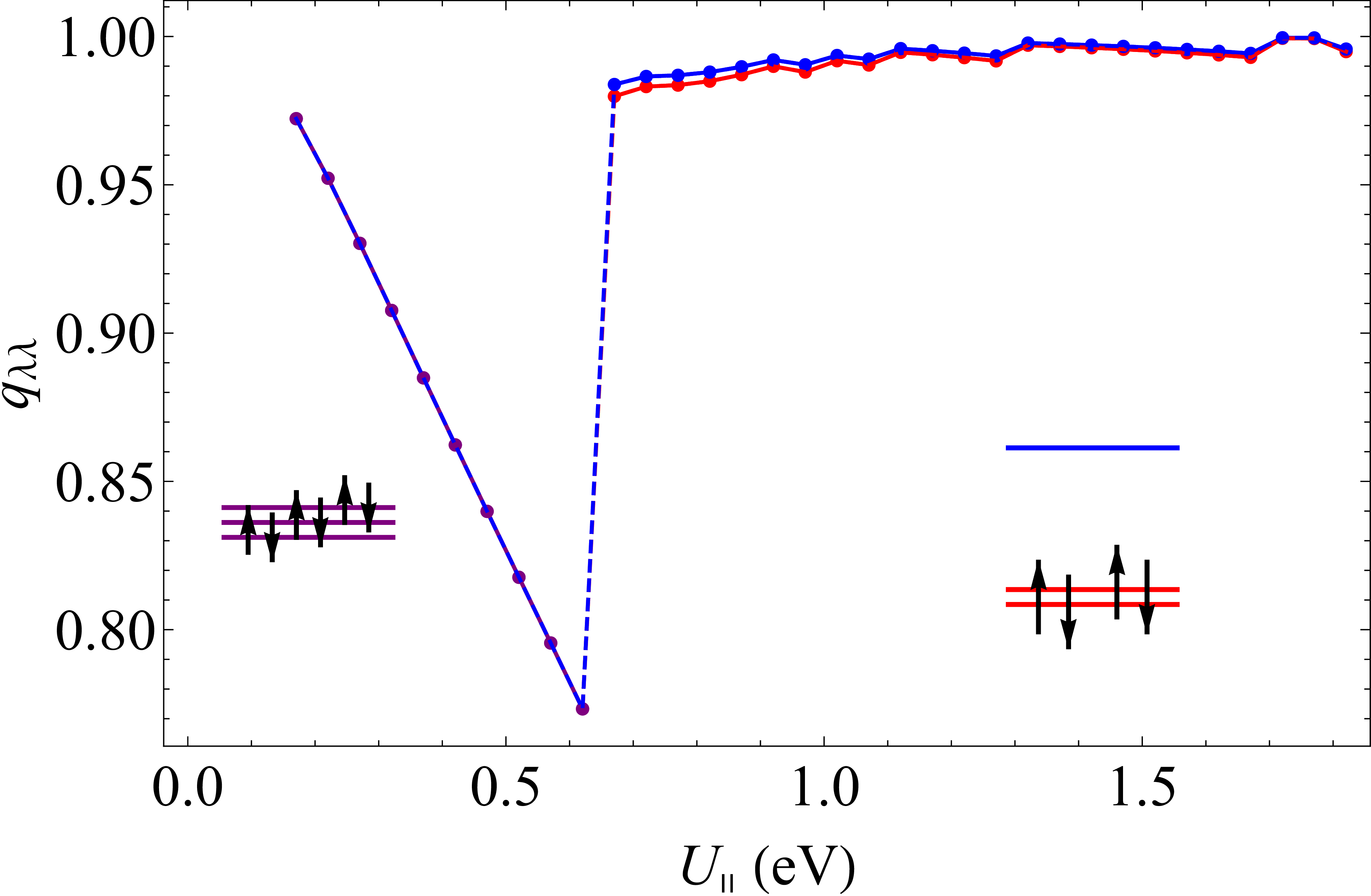}
&&
\includegraphics[width=7cm, bb= 0 0 3931 2625]{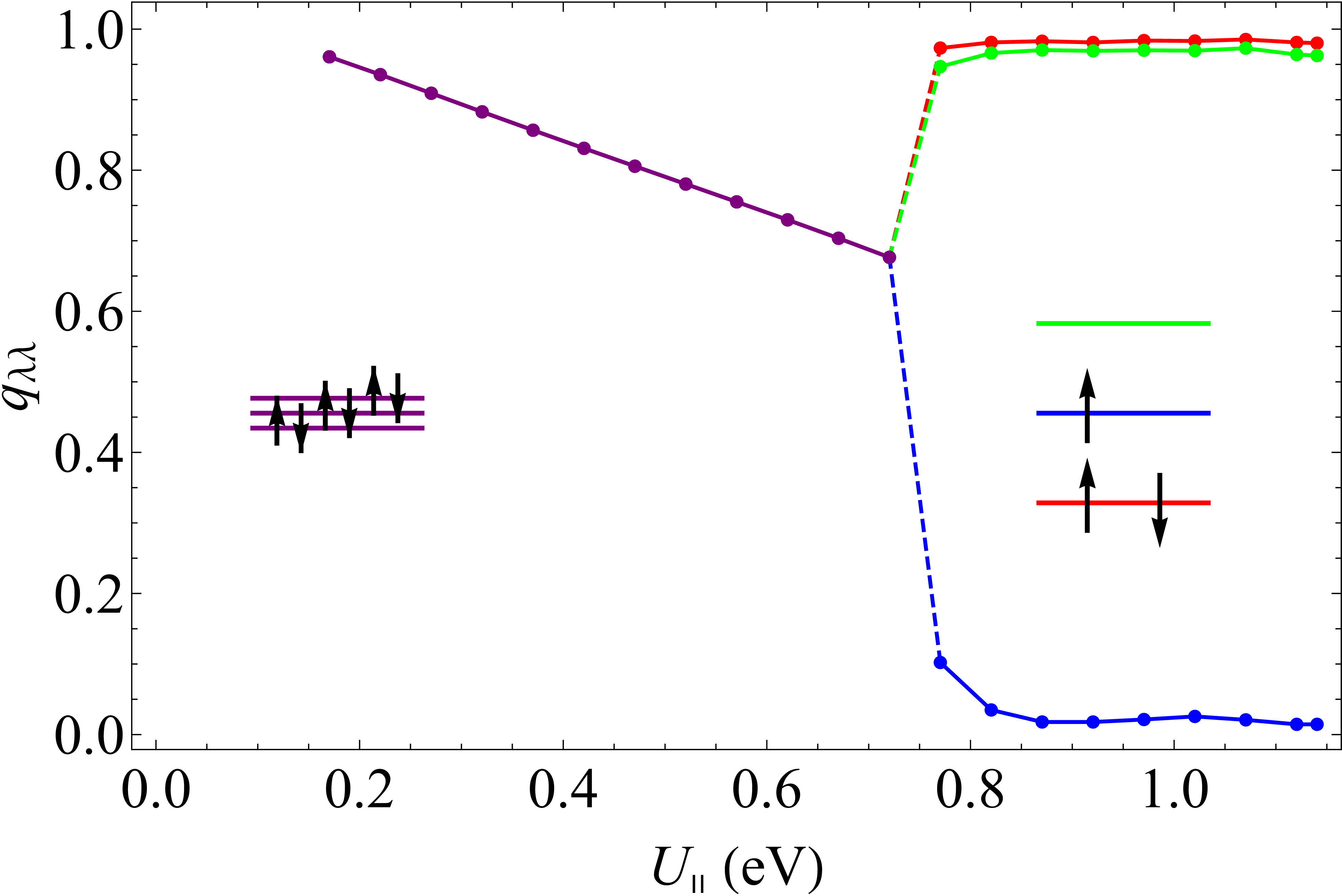}
\end{tabular}
\end{center}
\caption{
(A) Occupation numbers per electron spin of LUMO orbitals $n_{\lambda}$ 
and (C) Gutzwiller reduction factors in the corresponding bands for a model A$_4$C$_{60}$ with cubic band dispersion (see the text) 
subject to static JT interaction as function of $U_\parallel$. 
(B) and (D): the same as (A) and (C), respectively, for fcc A$_3$C$_{60}$. 
}
\label{Fig2}
\end{figure*}

The character of the electronic phase differs drastically in the two domains of $U_{\parallel}$.
The difference is clearly seen in the electron population in the LUMO orbitals $n_\lambda$ 
and the Gutzwiller's reduction factor $q_{\lambda \lambda}$. 
The evolution of the population $n_\lambda$ with respect to $U_\parallel$ (Fig. \ref{Fig2}(A)) 
shows that for $U_{\parallel} < U_{c}$ the phase corresponds to equally populated LUMO bands.
This equally populated phase gradually becomes strongly correlated with increasing $U_{\parallel}$, which is 
testified by the accompanying decrease of the Gutzwiller's reduction factors for these bands (Fig. \ref{Fig2}(C)). 
On the contrary, for $U_{\parallel} > U_{c}$, it exhibits orbital disproportionation of electronic density among the LUMO orbitals 
(Fig. \ref{Fig2}(A)) with a sudden jump of the Gutzwiller factor (Fig. \ref{Fig2}(C)). 

The existence of the two kinds of phases with and without the JT deformation is explained 
by the competition between the band energy $\langle \hat{H}_\text{t}\rangle$ and the JT stabilization energy 
in the presence of the strong electron repulsion $U_\parallel$.
The former stabilizes the system the most when the splitting of the orbital is absent, 
while the JT effect does by lowering the occupied orbitals. 
On the other hand, the bielectronic energy is reduced by the quenching of the charge fluctuation (localization of the electrons), 
which results in the decrease of the band energy and the relative enhancement of the JT stabilization. 
Therefore, when $U_\parallel$ is small ($U_\parallel < U_c$), the homogeneous (with equal orbital populations) band state 
is favored and the JT distortion is quenched.
With the increase of $U_\parallel$ over $U_c$, the band energy is reduced 
to the extent that the JT stabilization on C$_{60}$ sites is favored, resulting in orbitally disproportionated ground state.

We note that these results are general, which neither depends on the form of the JT distortion on sites 
nor on the uniformity of these distortions, which can also be dynamical as in A$_3$C$_{60}$ \cite{Iwahara2015} ({\em vide infra}).

\section{Band insulating state in the presence of strong electron repulsion}
\label{Sec:BI}
\begin{figure*}[bt]
\begin{center}
\begin{tabular}{lll}
(A) &~~& (B) \\ 
\includegraphics[width=7cm, bb= 0 0 4758 3050]{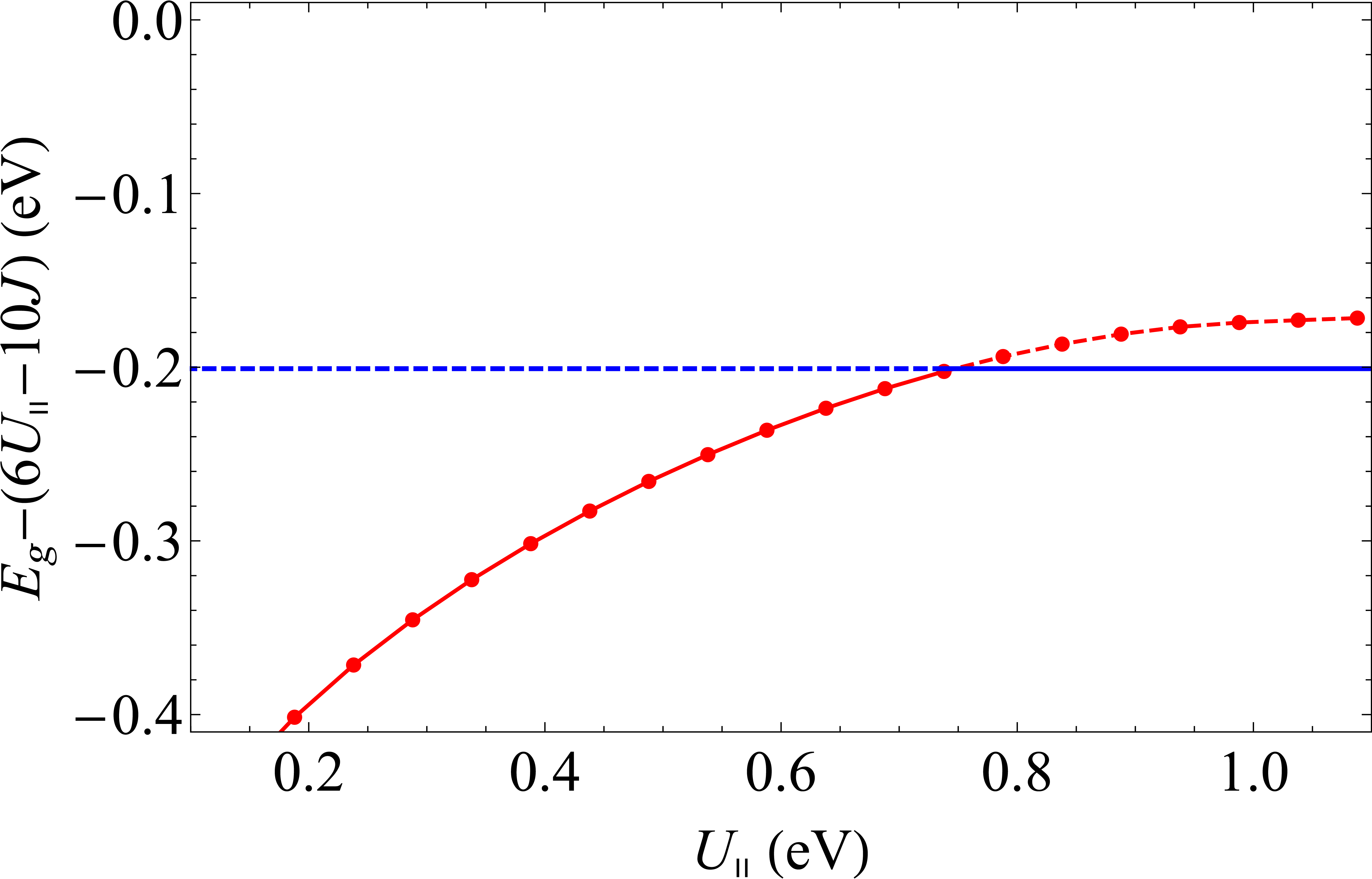}
&&
\includegraphics[width=7cm, bb= 0 0 3965 2542]{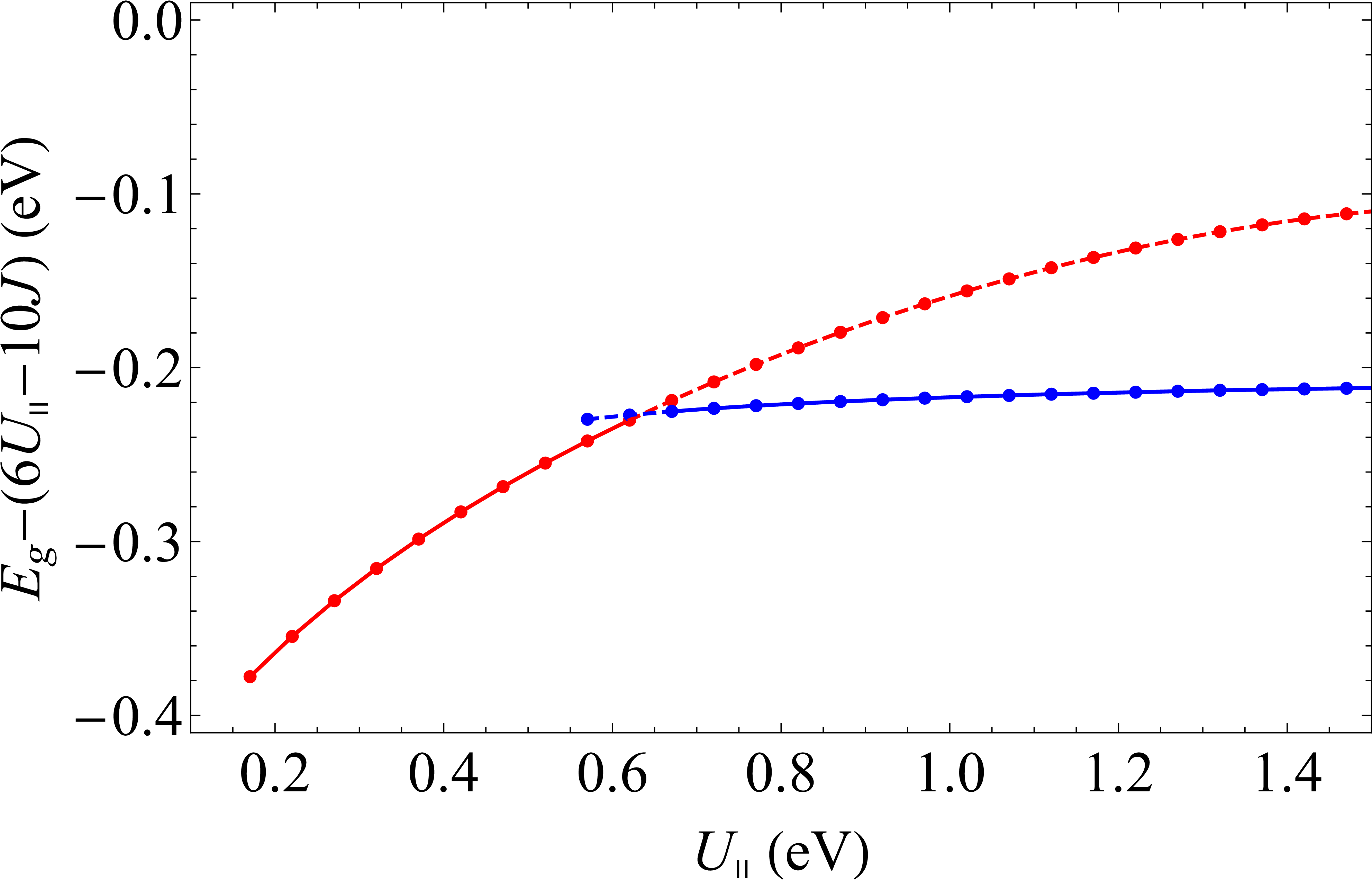}
\end{tabular}
\end{center}
\caption{
(A) Total energy of the ground electronic phase of A$_4$C$_{60}$ with cubic band dispersion 
and suppressed interband electron transfer 
($t^{\Delta \mathbf{m}}_{\lambda \lambda'} = \delta_{\lambda \lambda'} t^{\Delta \mathbf{m}}_{\lambda \lambda}$)
as function of $U_\parallel$. 
The red and the blue lines indicate the correlated band solution ($q=0$) and 
band insulating solution with JT splitting, respectively,
and the solid and dashed lines indicate the ground and the excited states, respectively, for each $U_\parallel$.
The bielectronic energy ($6U_\parallel - 10 J$) is subtracted from $E_g$.
(B) The same for the model of A$_4$C$_{60}$ 
with full transfer Hamiltonian used in Eq. (\ref{Eq_1}).
}
\label{Fig3}
\end{figure*}

To better understand the physics of the obtained orbitally disproportionated electronic phase,
first consider a simplified model for $\hat{H}_t$ which includes only the diagonal electron transfers after orbital indices,
$t^{\Delta \mathbf{m}}_{\lambda \lambda'} = \delta_{\lambda \lambda'} t^{\Delta \mathbf{m}}_{\lambda \lambda}$
(a widely used approximation for the study of multiorbital correlation effects \cite{Gunnarsson1996, Gunnarsson1997, Kita2011}). 
Figure \ref{Fig3}(A) shows the total energies for the two phases with and without JT distortion
in function of $U_{\parallel}$. 
We see again an evolution of the ground state with the stabilization of orbitally disproportionated electronic phase 
in the large $U_{\parallel}$ domain.
We find this behaviour pretty similar to the case when the full $\hat{H}_{\text{t}}$ for fcc lattice 
is considered (Fig. \ref{Fig3}(B)). 
Owing to the simplification, we can fully identify the orbitally disproportionated phase because have its exact solution. 
Indeed, in terms of band solutions $\hat{a}_{{\mathbf{k}}\alpha\sigma}^{\dagger}|0\rangle 
=1/\sqrt{N}\sum_{{\mathbf{m}}}\text{e}^{i{\mathbf{k}\cdot\mathbf{m}}}\hat{c}_{\mathbf{m}\alpha\sigma}^{\dagger}|0\rangle$, 
where $N$ is the number of sites, we obtain for the orbitally disproportionated phase (see Supplementary Material):   
\begin{equation}
|\Phi_0\rangle = 
 \prod_{{\mathbf{k}}\sigma}^{\text{all}} \hat{a}_{\mathbf{k}x\sigma}^{\dagger} \hat{a}_{\mathbf{k}y\sigma}^{\dagger} |0\rangle ,
\label{Eq_2}
\end{equation}
i.e., a pure band state with occupied $x$ and $y$ and empty $z$ band. In the case of a JT distortion different from the $h_g\theta$ type, 
the solution will be identical to Eq. (\ref{Eq_2}) but involving band orbitals which are linear combinations of $x$, $y$ and $z$ orbitals. 
The solution $\Phi_0$ is exact in the whole domain of $U_{\parallel}$. 
However, due to its fully disproportionated character, always corresponding to the orbital populations (2,2,0), 
it becomes ground state, i.e., intersects the correlated homogeneous solution (Fig. \ref{Fig3}(A)), 
only under the opening of the gap between occupied degenerate orbitals $x,y$ and the empty orbital $z$. 
This means that the orbitally disproportionated phase in Fig. \ref{Fig3}(A)  is nothing but conventional band insulator. 

The obtained result is not specific to the simplified model.
In the case of full $\hat{H}_{\text{t}}$ (Fig. \ref{Fig3}(B)),
the orbitally disproportionated state differs only slightly from $\Phi_0$ in Eq.  (\ref{Eq_2}), 
which is seen from the population of the orbital components of the LUMO band 
$n_\lambda$ that are close to (2,2,0), Fig. \ref{Fig2}(A), 
and the jump of the Gutzwiller factor to its uncorrelated value 1, Fig. \ref{Fig2}(C).
Thus, we encounter here a counterintuitive situation: 
with the increase of the electron repulsion on sites, the system passes 
from a strongly correlated metal to a uncorrelated band insulator.  

To get further insight into the correlated metal to band insulator transition, 
we compare the electronic state of A$_4$C$_{60}$ with that of the correlated A$_3$C$_{60}$ 
which turns into MH insulator for large $U_\parallel$.
In both fullerides, the transition from the orbitally degenerate phase to the disproportionated phase is observed 
with the increase of $U_\parallel$, however, the nature of the latter phases is significantly different.
Because orbital disproportionation is indissolubly linked to JT distortions on fullerene sites, either static or dynamic, 
the LUMO band in A$_3$C$_{60}$ will be split in three orbital subbands. 
Figures \ref{Fig2}(B) and \ref{Fig2}(D) show that the lowest orbital subband in A$_3$C$_{60}$ becomes fully 
occupied and practically uncorrelated ($q_{\lambda \lambda} \approx 1$) with increase of $U_{\parallel}$ 
in very close analogy with the behavior of the two lowest subbands in 
A$_4$C$_{60}$ (Fig. \ref{Fig2}(C)). 
At the same time the electron correlation in the middle half-occupied subband gradually increases implying 
that the MH transition basically occurs in this subband \cite{Iwahara2015}.
Indeed, the bielectronic energy is reduced by quenching the charge fluctuations in the half-filled middle subband. 
This is seen as the decrease of the Gutzwiller's factor with the increase of $U_\parallel$ (Fig. \ref{Fig2}(D)),
testifying about suppression of the intersite electron hopping.
On the contrary, the doubly occupied orbitals are not subject to electron correlation
(Gutzwiller's factor becomes close to 1, Fig. \ref{Fig2}(D)).
In the case of A$_4$C$_{60}$, the LUMO orbitals split into two doubly filled orbitals and 
non-degenerate empty orbital by the JT interaction (see the inset of Fig. \ref{Fig2}(A)).
The fully occupied orbitals are similar in nature to those of A$_3$C$_{60}$, being basically uncorrelated, 
the same for the empty orbital
(all Gutzwiller's factors are close to 1, Fig. \ref{Fig2}(C)).

\begin{table}[tb]
\caption{
The criterion for correlated metal to band insulator transition in three-fold degenerate band system
with four electrons per site. 
\footnote{
$+/-$ stands for presence/absence, 
$E_\text{JT}$ is the JT stabilization energy for C$_{60}^{4-}$, 
$\Delta_0$ is the (non-JT) crystal-field splitting of the $t_{1u}$ LUMO shell on one fullerene site,
$\hslash\bar{\omega}/2$ is the energy gain due to the JT dynamics per dimension of the trough,
and $0 \le \mu \le 1$.
The Hund's rule energy $3J$ will be slightly modified by taking into account the multiplet structure 
due to the presence of two low-spin terms in C$_{60}^{4-}$.
}
}
\label{TableI}
\begin{ruledtabular}
\begin{center}
\begin{tabular}{ccccc}
&Intrinsic & Static & Dynamical & Band \\
&orbital splitting       & JTE    & JTE       & insulator \\
\hline
\multirow{3}{*}{(A)}
&
\multirow{3}{*}{
\includegraphics[width=0.8cm, bb= 0 0 417 104]{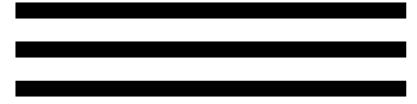}
}
           & $-$ & $-$ & Never \\
         & & $+$ & $-$ & $E_\text{JT} > 3J$ \\
         & & $+$ & $+$ & $E_\text{JT} + \hslash \bar{\omega} > 3J$ \\
\hline
\multirow{3}{*}{(B)}
&
\multirow{3}{*}{
\includegraphics[width=0.8cm, bb= 0 0 417 437]{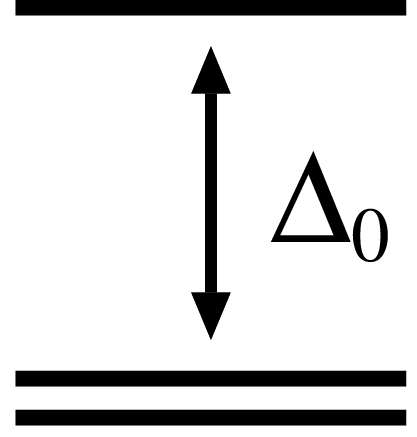}
}
          & $-$ & $-$ & $\Delta_0 > 3J$ \\
        & & $+$ & $-$ & $\Delta_0 + E_\text{JT} > 3J$ \\
        & & $+$ & $+$ & $\Delta_0 + E_\text{JT} + \frac{1}{2} \hslash \bar{\omega} > 3J$ \\
\hline
\multirow{3}{*}{(C)}
&
\multirow{3}{*}{
\includegraphics[width=0.8cm, bb= 0 0 417 444]{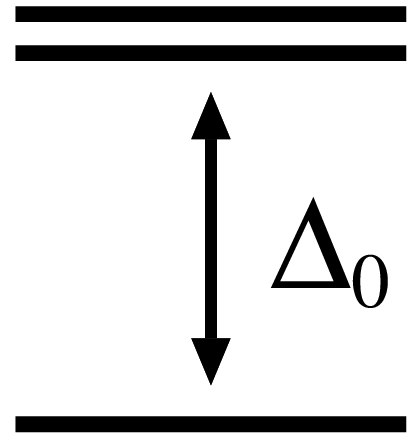}
}
          & $-$ & $-$ & Never \\
        & & $+$ & $-$ & $E_\text{JT} > 3J$ \\
        & & $+$ & $+$ & $E_\text{JT} + \frac{1}{2} \hslash \bar{\omega} > 3J$ \\
\hline
\multirow{3}{*}{(D)}
&
\multirow{3}{*}{
\includegraphics[width=1.4cm, bb= 0 0 417 267]{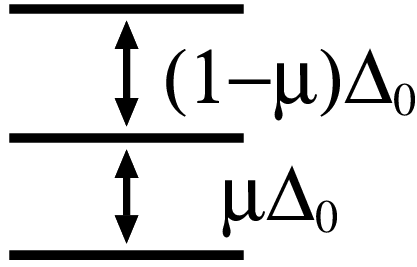}
}
          & $-$ & $-$ & $(1-\mu) \Delta_0 > 3J$ \\
        & & $+$ & $-$ & $(1-\mu) \Delta_0 + E_\text{JT} > 3J$ \\
        & & $+$ & $+$ & $(1-\mu) \Delta_0 + E_\text{JT} > 3J$ \\
\end{tabular}
\end{center}
\end{ruledtabular}
\end{table}

\section{Condition for the stabilization of orbitally disproportionated phase}
\label{Sec:Condition}
The necessary condition for achieving the band insulating state 
is that in the atomic limit of large $U_{\parallel}$, the orbitally disproportionated molecular state ($S=0$)
has lower energy than the homogeneous $S=1$ Hund state on each C$_{60}$. 
Consider the $t_{1u}$ orbital shell of one single fullerene site. 
Due to the Hund's rule coupling, the high-spin configurations ($S=1$), e.g., (2,1,1),
are stabilized by $3J$ with respect to the low-spin configurations ($S=0$), e.g., (2,2,0).
The high-spin (Hund) state always contains half-filled orbitals and leads, therefore, to MH insulator in the limit of large $U_\parallel$.
On the other hand, in the presence of a relatively strong 
static JT effect, the low-spin state is stabilized by $E_\text{JT} = 4E_\text{JT}^{(1)}$, 
where $E_{\text{JT}}^{(1)} = \hslash \omega g^2/2$ is the JT stabilization energy in C$_{60}^{-}$ \cite{Auerbach1994, OBrien1996}.
Thus, the low-spin state, and, consequently, the band insulating state, 
is realized as the ground state when the condition $E_\text{JT} > 3J$ is fulfilled.
With the estimate $E_{\text{JT}}^{(1)}=$ 50 meV and $J=$ 44 meV \cite{Iwahara2010, Iwahara2013},
we conclude that all A$_4$C$_{60}$ with hypothetical cubic structure will be band insulators in the static JT limit 
at sufficiently large $U_{\parallel}$. 

This condition is modified when there is an intrinsic orbital gap $\Delta_0$ at fullerene sites 
which arises due to the lowering of the symmetry of the crystal field (CF) in non-cubic fullerides (Table \ref{TableI}).
Band structure calculations of A$_4$C$_{60}$ with body centered tetragonal (bct) lattice show
that the low-symmetry CF is weak and does not admix the excited electronic states on fullerene sites.
Accordingly, the strength of the JT coupling is not modified by this CF splitting.
When one of the $t_{1u}$ orbitals is destabilized by the CF splitting $\Delta_0$ 
(Table \ref{TableI} (B)), the Hund configuration (2,1,1), with $S=1$, is also destabilized by $\Delta_0$,
whereas the energy of the low-spin configuration (2,2,0), 
with $S=0$, remains unchanged because the destabilized orbital is not populated ($n=0$).
The orbitally disproportionated state becomes the ground one when $E_\text{JT} + \Delta_0 > 3J$,
which means that the low-symmetry CF splitting enhances the tendency toward disproportionation. 
Moreover, if the CF splitting $\Delta_0$ is larger than the Hund's rule energy $3J$, 
the system becomes band insulator for sufficiently large $U_\parallel$ even in the absence of the JT effect
($E_\text{JT} = 0$).

On the contrary, if two $t_{1u}$ orbitals are equally destabilized by $\Delta_0$ (Table \ref{TableI} (C)), 
both the high-spin and the low-spin configurations are destabilized by $2\Delta_0$, thus the system does never become band insulator
only due to CF splitting. 
The band insulator is achieved in this case only when the JT stabilization in the low-spin state is stronger than the Hund energy $3J$, 
which results in the same criterion as for the degenerate case (A).
We stress that the amplitude of the CF splitting does not play a role in this case. 
It only plays a role when the destabilizations of the low- and high-spin configurations are different,
such as in the case of the second scenario (B) or the last one (D) corresponding to complete CF lift of degeneracy.
In the latter case, on the argument given above, only the CF splitting between the highest two orbitals adds to the criterion,
which looks now as intermediate ($0 < 1-\mu < 1$, see Table \ref{TableI}(D)) to the previous scenarios, (B) and (C).

According to the tight-binding simulations of the DFT LUMO band (Fig. \ref{Fig5}(A)), 
the pattern of the orbital splitting for the bct K$_4$C$_{60}$ corresponds to 
the third scenario of the CF splitting (Table \ref{TableI}(C)) with a gap $\Delta_0$ of ca 130 meV.
Given a similar lattice structure, the same situation is expected also for Rb$_4$C$_{60}$.
Therefore, according to the criterion in Table \ref{TableI}, no band insulating state can arise in these two fullerides,
unless the JT stabilization energy exceeds the Hund energy ($3J$). 
Following the estimations of $E_{\text{JT}}^{(1)}$ and $J$ (see above), 
we conclude that the uncorrelated band insulating phase is stabilized in A$_4$C$_{60}$ with A = K, Rb, 
in agreement with experiment. 
In body centered orthorhombic (bco) Cs$_4$C$_{60}$,
the low-symmetric CF will completely lift the degeneracy of the $t_{1u}$ orbitals, leading to a scenario (D) in Table \ref{TableI}. 
The splitting between the highest and the middle $t_{1u}$ orbitals will enhance the tendency towards the stabilization of
the band insulating state, according to the criterion in Table \ref{TableI}.

Finally, we consider the effect of the JT dynamics on the stabilization of the orbitally disproportionated phase.
In the cubic A$_4$C$_{60}$, 
due to a perfect disproportionation (2,2,0) of the occupation of orbital subbands, 
the dynamical JT effect on the fullerene sites will be unhindered by hybridization of orbitals between sites 
pretty much as in metallic A$_3$C$_{60}$ close to MH transition \cite{Iwahara2015}. 
The pseudorotation
of JT deformations in the trough of the ground adiabatic potential surface of fullerene anion gives a gain in nuclear kinetic energy of 
$\hslash\bar{\omega}/2 \approx 30$ meV per dimension of the trough \cite{Iwahara2013}.
The gain amounts to $\hslash\bar{\omega}$ in the case of two-dimensional trough in C$_{60}^{4-}$ \cite{Auerbach1994, OBrien1996}. 
This will enhance the criterion for band insulator by $\hslash\bar{\omega}$ in the case of cubic lattice (Table \ref{TableI}). 
For relatively large intrinsic CF gap, $\Delta_0 > \hslash\bar{\omega}/2$, one of the rotational degrees of freedom 
in the trough will be quenched and the JT dynamics will reduce to a one-dimensional 
pseudorotation of JT deformations entraining only the two degenerate orbitals in the (B) and (C)
scenarios of splitting shown in Table \ref{TableI}. 
This is apparently the case of bct K$_4$C$_{60}$ and Rb$_4$C$_{60}$ at ambient pressure.
In the case of last scenario (D) of CF splitting, the JT pseudorotational dynamics will be completely quenched if the separations 
between the three orbitals exceed much $\hslash \bar{\omega}/2$.
Whether this is the case of Cs$_4$C$_{60}$ with a relevant bco lattice, remains to be
answered by a DFT based analysis similar to one done here for K$_4$C$_{60}$ (Figs. \ref{Fig5}(A), (C)).

Another ingredient defining the transition from the correlated metal to band insulator is the bielectronic interaction $U_\parallel$.
The value of $U_{\parallel}$ at which the band insulating state is stabilized (the crossing point of the two phases in Fig. \ref{Fig3}) 
depends on the relation between the band energy in the homogeneous correlated metal phase $\langle\hat{H}_{\text{t}}\rangle$ 
and the gain of intrasite energy due to disproportionated orbital occupations (static and dynamic JT stabilization energies). 
The calculations (Fig. \ref{Fig3}) show that in the cubic model of 
A$_4$C$_{60}$, the band insulating state arises already 
at modest values of $U_{\parallel}$, which means that it is always achieved in these fullerides 
(cf. experimental Hubbard $U \approx$ 0.4-0.6 eV for K$_3$C$_{60}$ \cite{Bruhwiller1993, Macovez2011}). 
Since the necessary conditions for the cubic and bct A$_4$C$_{60}$ are the same (Table \ref{TableI}), 
the band insulating state seems to be well achieved in the bct K$_4$C$_{60}$ and Rb$_4$C$_{60}$.
The stabilization of the band insulating state in the bco Cs$_4$C$_{60}$ seems to 
be facilitated by a larger $U_\parallel$ expected due to the larger distance between C$_{60}$ sites. 
This is in line with the experimental observation of insulating nonmagnetic state in all A$_4$C$_{60}$ at ambient pressure 
\cite{Kiefl1992, Murphy1992, Kosaka1993}. 

We want to emphasize that the intrinsic CF splitting of the $t_{1u}$ LUMO orbitals on C$_{60}$ sites in fullerides does not render them
automatically band insulators. 
Thus, the DFT calculations of K$_4$C$_{60}$ (Figs. \ref{Fig5}(A) and \ref{Fig5}(C)) do not give a
band insulator but rather a metal despite the intrinsic CF splitting of 130 meV. 
The same situation is realized in Cs$_4$C$_{60}$ and any other fulleride in which the intrinsic CF splitting is 
significantly smaller than the uncorrelated bandwidth. 
The band insulating state (Figs. \ref{Fig5}(B) and \ref{Fig5}(D)) only arises due to JT distortions on fullerene
sites and due to the effects of electron repulsion in the $t_{1u}$ shell reducing much the band energy 
of the homogeneous metallic state. 

\begin{figure}[bt]
\begin{center}
 \includegraphics[width=8cm, bb= 0 0 3931 2618]{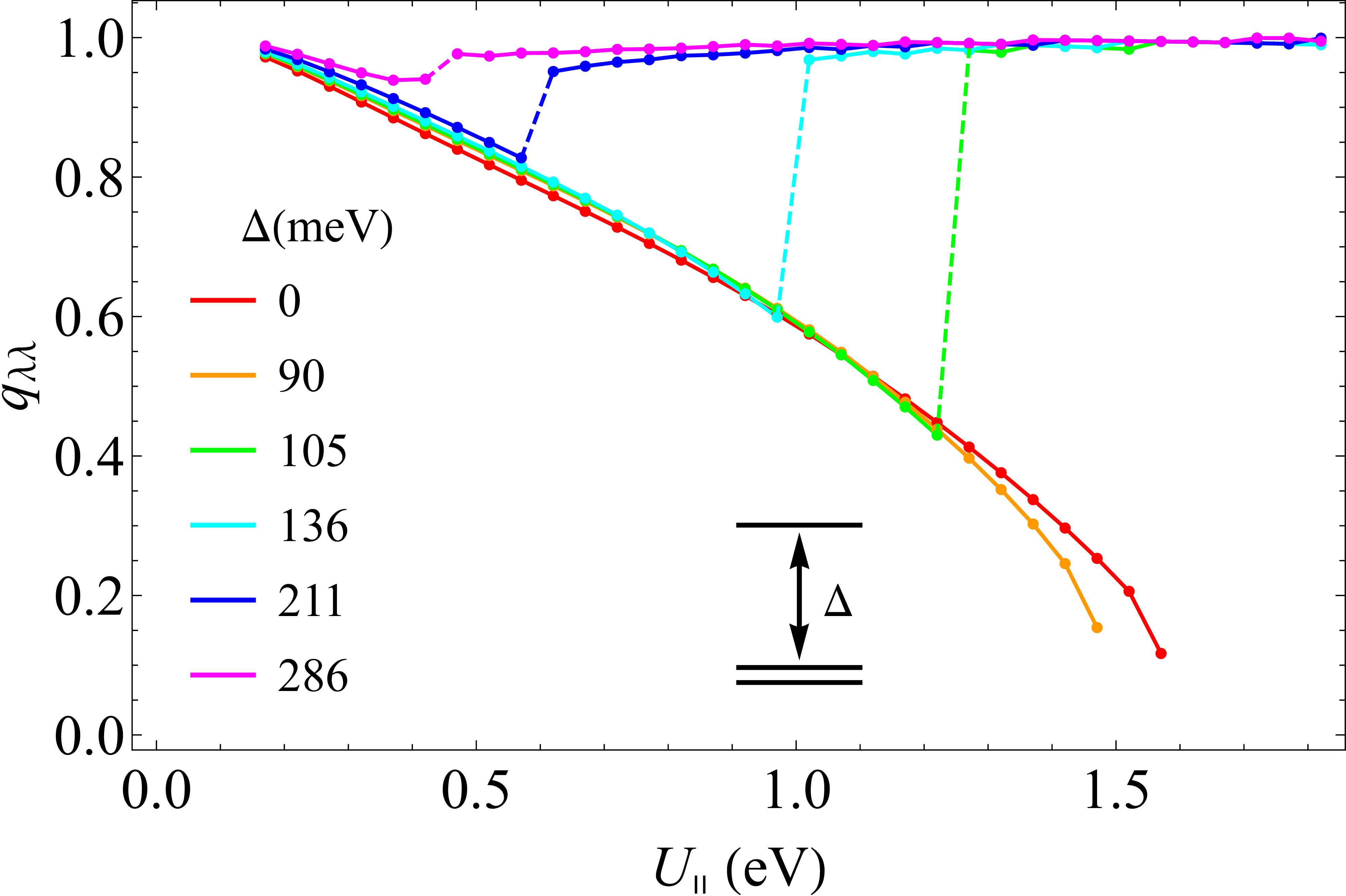}
\end{center}
\caption{
Evolution of the the Gutzwiller reduction factors $q_{\lambda \lambda}$ for A$_4$C$_{60}$ within the cubic
model used in Fig. \ref{Fig1} in function of $U_\parallel$ for different orbital gaps
$\Delta$, which are sums of JT and CF splittings (the former is considered arbitrary now).
The monotonously decreasing line corresponds to a correlated metal, which for $\Delta < \Delta_c$ 
($\Delta_c \approx 100$) evolves into a MH insulator.
The jumps to $q_{\lambda \lambda} \approx 1$ for values $\Delta > \Delta_c$ correspond to onsets of band insulator.  
}
\label{Fig4}
\end{figure}

Generalizing, the band insulating state will be achieved
at any value of the gap between the highest and the middle LUMO orbitals 
$\Delta$ (a sum of CF and JT splittings) at C$_{60}$ sites which fulfills
the necessary condition in Table \ref{TableI}. 
The only difference is that smaller $\Delta$ will require larger $U_\parallel$ for achieving the intersection 
with the homogeneous correlated metal phase (Fig. \ref{Fig4}).
One should note that the band insulating state arises not only three-orbital systems like fullerides,
but also in other orbitally degenerate systems with even numbers of electrons per site 
when both $\Delta$ and $U_\parallel$ are sufficiently large.
Thus the scenario (B) without JT effect in Table \ref{TableI} was considered for a 1/3-filled three-orbital model
with infinite-dimensional Bethe lattice \cite{Kita2011}.

\section{Universality of orbital disproportionation in fullerides}
\label{Sec:Universality}
Given the established orbital disproportionation of the LUMO electronic density in A$_3$C$_{60}$ \cite{Iwahara2013, Iwahara2015}, 
its persistence in A$_4$C$_{60}$ found in the present work makes the orbital disproportionation a universal 
feature of electronic phases in alkali-doped fullerides.
Indeed, the same electronic phase is expected also for A$_2$C$_{60}$ fullerides \cite{Murphy1992, Brouet2001},
which are described by essentially the same interactions as A$_4$C$_{60}$. 
The only difference will be the inversion of the intrinsic CF 
and JT orbital splittings on the fullerene sites.

The existence of the orbital disproportionation in fullerides is imprinted on their basic electronic properties.
As discussed in Sec. \ref{Sec:BI} and Ref. \cite{Iwahara2015}, 
in the disproportionated phase the orbital degeneracy is lifted and the electron correlation 
develops in the middle subband, whereas it does not play a role in other subbands. 
Therefore, the MH transition also mainly develops in the middle subband \cite{Iwahara2015}, and hence, 
one has no ground whatsoever to claim strong effects 
of orbital degeneracy on the MH transition in these materials as was done repeatedly in the past 
\cite{Lu1994,Gunnarsson1996,Koch1999II}. 
Another important manifestation of the orbital disproportionation is the similar JT dynamics corresponding to independent 
pseudorotation of JT deformations on different fullerene sites in both MH phase \cite{Iwahara2013} 
and strongly correlated metallic phase \cite{Iwahara2015} of A$_3$C$_{60}$. 
This has recently found a firm experimental confirmation in the equivalence of IR spectra of the corresponding materials
\cite{Zadik2015}.

In A$_4$C$_{60}$, the experimental evidence for the (2,2,0) orbital disproportionated phase comes, first of all, from the observed 
non-magnetic insulating ground state.
Moreover, as implied by the intersection picture of the two ground phases (Fig. \ref{Fig3}), 
the correlated metal to band insulator transition could be observed by the decrease of the bielectronic interaction $U_\parallel$ 
with respect to the band energy.
This seems to be realized as the metal-insulator transition in Rb$_4$C$_{60}$ under pressure \cite{Kerkoud1996},
where the electron transfer (band energy) is enhanced by the decrease of the distance between the sites 
and $U_\parallel$ is concomitantly reduced by the enhanced screening.

Further evidence for the orbitally disproportionated phase comes from spectroscopy. 
In the case of static JT distortions of $h_g\theta$ type on fullerene sites, 
the single-particle excitations are exactly described by the uncorrelated band solutions,
$|\Phi_{z{\mathbf{k}}\sigma}^e\rangle =\hat{a}_{z{\mathbf{k}}\sigma}^{\dagger} |\Phi_0\rangle$ for electron and 
$|\Phi_{\alpha{\mathbf{k}}\sigma}^h\rangle =\hat{a}_{\alpha{\mathbf{k}}\sigma} |\Phi_0\rangle$, 
$\alpha =x,y$, for hole quasiparticles, respectively (see the Supplementary Materials). 
Figure \ref{Fig5} shows that the dispersion of electron- and hole-like excitation basically corresponds 
to the decoupled $z$ and ($x,y$) bands due to practically suppressed hybridization between occupied 
and unoccupied LUMO orbitals when the band gap opens. 
The hole-like excitations (Fig. \ref{Fig5}(D)) show the density of states closely 
resembling the width and the shape of the LUMO feature in the photoemiossion spectrum \cite{DeSeta1995}.

\begin{figure*}[bt]
\begin{center}
\begin{tabular}{lll}
(A) &~~& (B) \\
\includegraphics[width=7cm, bb= 0 0 3965 2424]{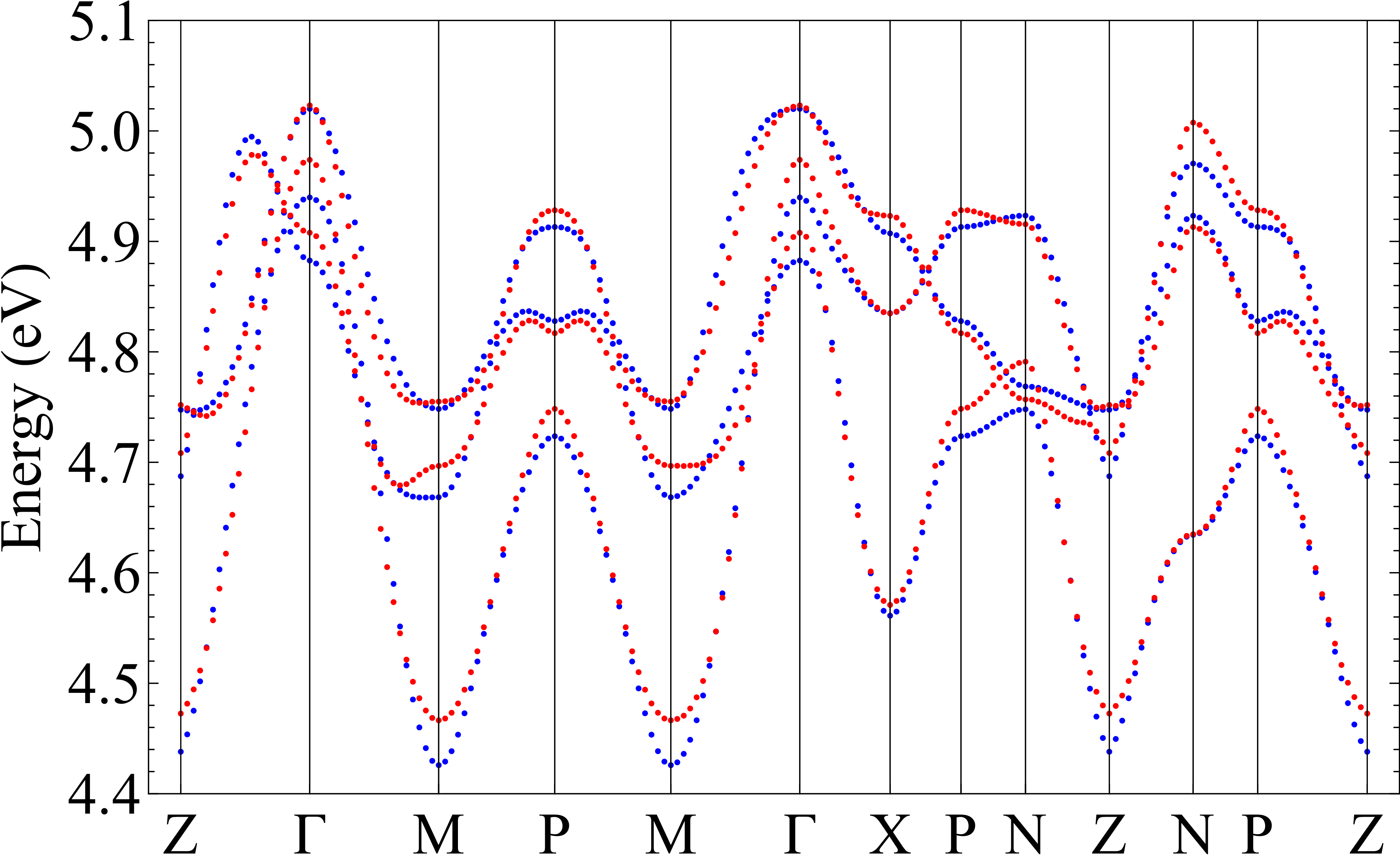}
&&
\includegraphics[width=7cm, bb= 0 0 3965 2368]{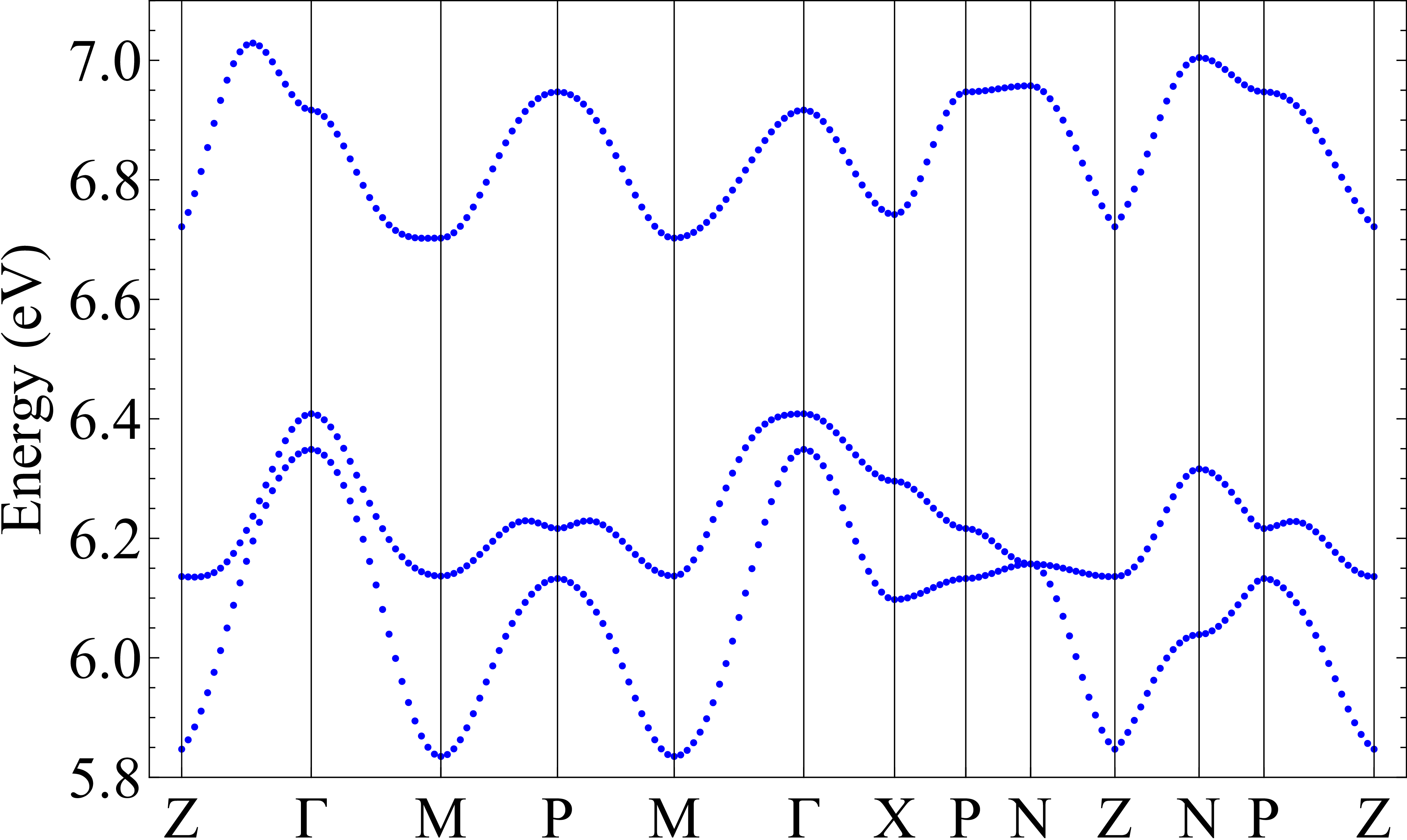}
\\
(C) && (D) \\
\includegraphics[width=7cm, bb= 0 0 3958 2625]{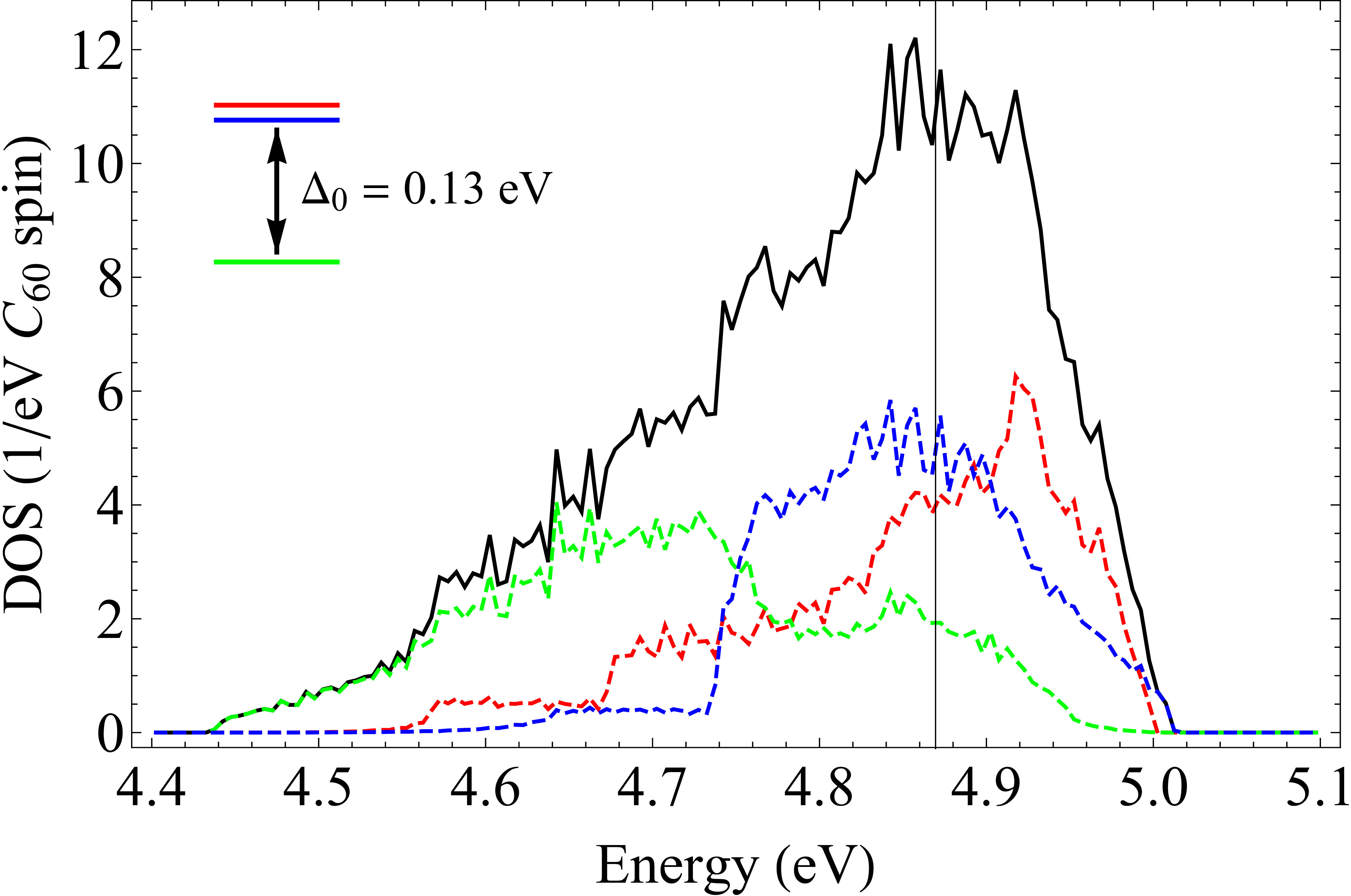}
&&
\includegraphics[width=7cm, bb= 0 0 3965 2687]{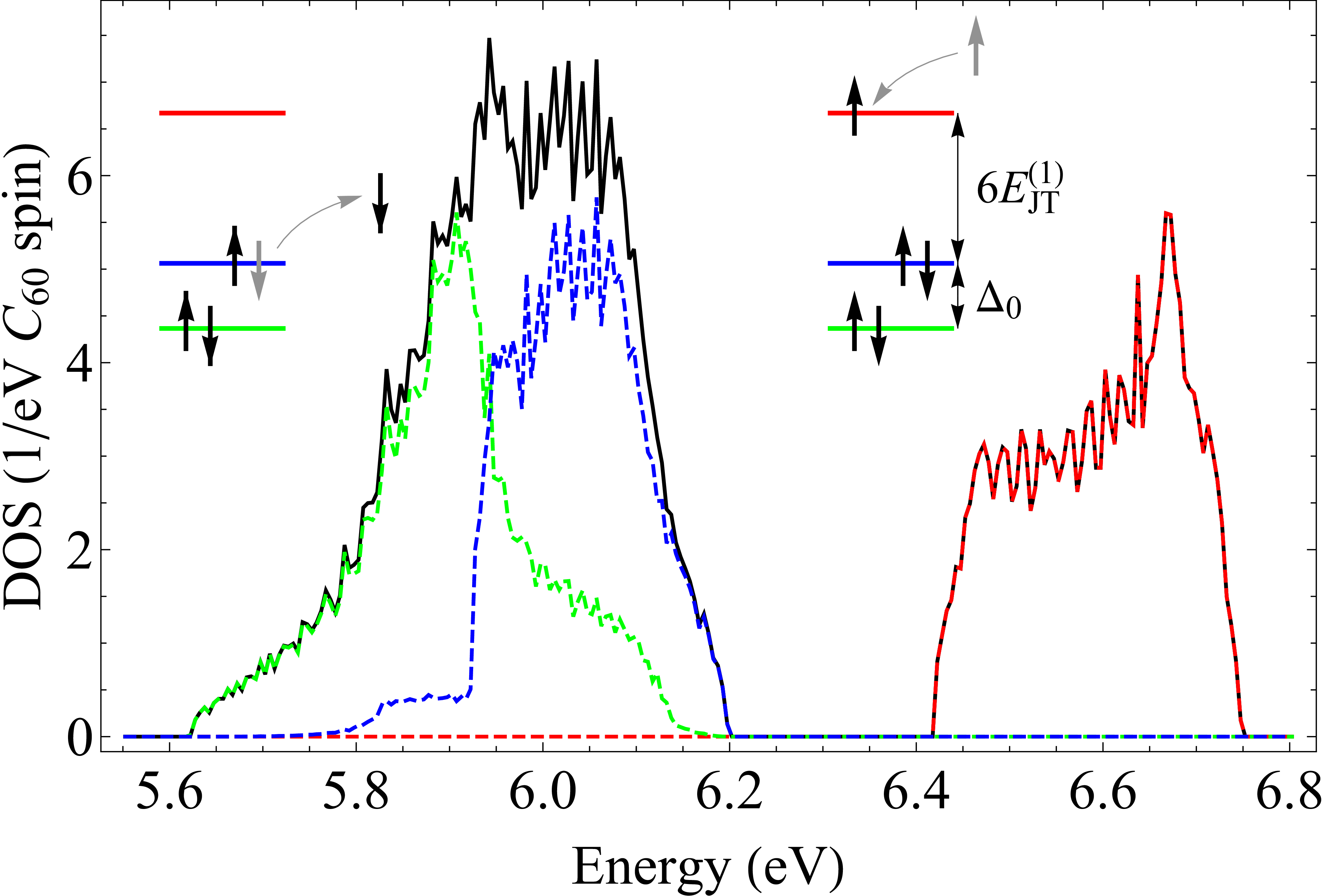}
\end{tabular}
\end{center}
\caption{
(A) LUMO band dispersion and (C) corresponding density of states of K$_4$C$_{60}$ calculated by DFT (GGA) for experimental structure. 
(B) Dispersion of single-particle excitations and (D) the corresponding density of states corresponding to the orbitally disproportionated 
ground state at $U_{\parallel}$= 0.5 eV and without hybridization between the occupied and empty band orbitals. 
The blue dots in (A) and (C) show the tight-binding simulation,
and red dots in (A) the DFT calculations.
The black line in (C) and (D) corresponds to a full DOS, while the colored lines the orbitally projected DOS.
}
\label{Fig5}
\end{figure*}

\section{Discussion}
\label{Sec:Discussion}
In this work, we investigated theoretically the ground electronic phase of A$_4$C$_{60}$ fullerides. 
It is found that the relatively strong electron repulsion on C$_{60}$ sites stabilizes the uncorrelated band insulating state in these materials. 
A particular conclusion of the present study is that the widely used term ``Jahn-Teller-Mott insulator'' 
\cite{Fabrizio1997, Knupfer1997, Brouet2004, Klupp2006}
is not appropriate here because it involves mutually excluding phenomena. 
A$_4$C$_{60}$ or any similar multiorbital system with even number of electrons per sites can be either a 
correlated metal with no JT distortions, high-spin (Hund)
MH insulator, or uncorrelated band insulator stabilized by static or dynamic JT distortions. 
We prove here that the latter is the case in 
the fullerides due to a weaker Hund's rule interaction compared to JT stabilization energy, 
which is ultimately due to relatively large radius of C$_{60}$. 
Similar situation should arise in other crystals with large unit cells with local orbital degeneracy, 
the first candidate being the molecular crystals of K$_4$ clusters \cite{Rao1985}. 
The present demonstration of the persistence of band insulating phase in A$_n$C$_{60}$ with even $n$ identifies the orbital 
disproportionation of the LUMO electronic density as a universal key feature of all alkali-doped fullerides, which undoubtly 
has a strong effect on their electronic properties.   
We would like to emphasize that the ultimate reason of orbital disproportionation in fullerides
is the existence of equilibrium Jahn-Teller distortions, static or dynamic, on fullerene sites.
These are always present in fullerides due to the crucial effect of electron correlation 
on the Jahn-Teller instability of C$_{60}^{n-}$ sites.

%


\clearpage
\begin{center}
\textbf{
Supplemental Materials\\
for
\\
``Orbital disproportionation of electronic density - a universal feature of alkali-doped fullerides''
}
\end{center}
\setcounter{equation}{0}
\setcounter{figure}{0}
\setcounter{table}{0}
\setcounter{section}{0}
\makeatletter
\renewcommand{\theequation}{S\arabic{equation}}
\renewcommand{\thefigure}{S\arabic{figure}}
\renewcommand{\thetable}{S\arabic{table}}
\renewcommand{\bibnumfmt}[1]{[S#1]}
\renewcommand{\citenumfont}[1]{S#1}

This Supplemental Materials include:
\begin{enumerate}
\item Jahn-Teller effect of isolated C$_{60}^{n-}$,
\item Explanation of the Gutzwiller's approach to the Jahn-Teller system, 
\item Ground state of fcc K$_3$C$_{60}$,
\item The band structure and transfer parameters of K$_4$C$_{60}$,
\item Band structure and DOS of K$_4$C$_{60}$ with and without interorbital hybridization,
\item Eigenstates of non-hybridized system.
\end{enumerate}

\section{Jahn-Teller effect of isolated C$_{60}^{n-}$}
\begin{figure}[bth]
\begin{center}
\includegraphics[height=4cm,bb=0 0 3070 3007]{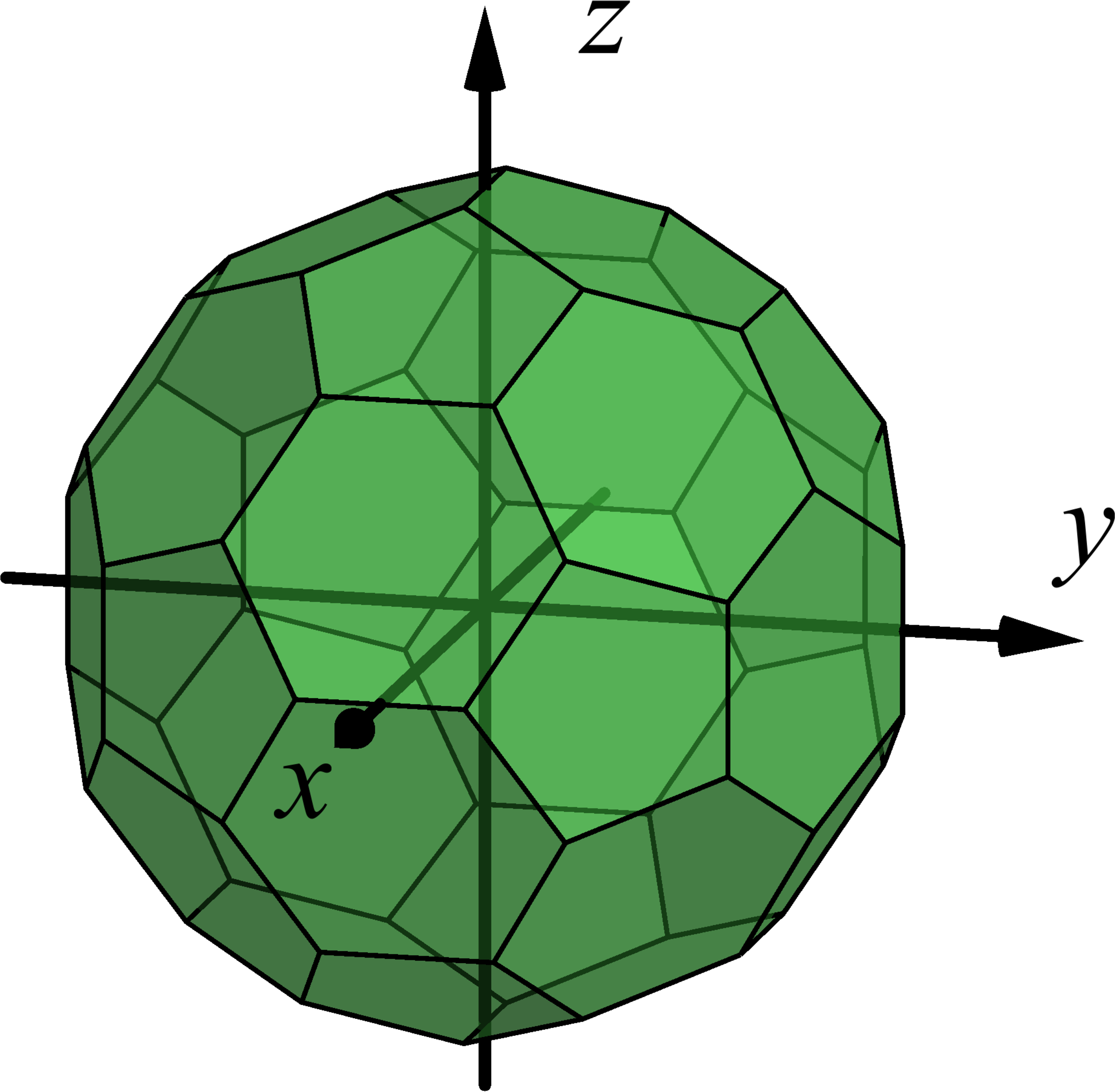}
\end{center}
\caption{Coordinate system of C$_{60}$.}
\label{Fig:C60}
\end{figure}

Isolated C$_{60}$ molecule ($I_h$ symmetry) has triply degenerate $t_{1u}$ LUMO level. 
The $t_{1u}$ orbital couples to two nondegenerate $a_g$ normal vibrational modes and 
eight five-fold degenerate $h_g$ normal vibrational modes. 
In this work, we omit the $a_g$ modes and use effective $h_g$ mode. 
The Jahn-Teller (JT) Hamiltonian of C$_{60}^{n-}$ $(n=1-5)$ is written as 
\begin{eqnarray}
 H_\text{JT} &=& \sum_{\gamma}\frac{\hslash \omega}{2} \left(p_\gamma^2 + q_\gamma^2\right) + \hslash \omega g 
 \sum_\sigma
 \left(\hat{c}_{x\sigma}^\dagger, \hat{c}_{y\sigma}^\dagger, \hat{c}_{z\sigma}^\dagger \right)
\nonumber\\
 &\times&
 \begin{pmatrix}
  \frac{1}{2}q_\theta - \frac{\sqrt{3}}{2} q_\epsilon & -\frac{\sqrt{3}}{2} q_\zeta & -\frac{\sqrt{3}}{2} q_\eta \\
 -\frac{\sqrt{3}}{2} q_\zeta & \frac{1}{2}q_\theta + \frac{\sqrt{3}}{2} q_\epsilon & -\frac{\sqrt{3}}{2} q_\xi \\ 
 -\frac{\sqrt{3}}{2} q_\eta & -\frac{\sqrt{3}}{2} q_\xi & -q_\theta 
 \end{pmatrix}
 \begin{pmatrix}
  \hat{c}_{x\sigma}\\ 
  \hat{c}_{y\sigma}\\
  \hat{c}_{z\sigma}
 \end{pmatrix},
\nonumber\\
\label{Eq:UJT}
\end{eqnarray}
where 
$\omega$ is the frequency, $g$ is the dimensionless vibronic coupling constant, 
$(q_\theta, q_\epsilon, q_\xi, q_\eta, q_\zeta)$ 
are the dimensionless mass-weighted normal vibrational coordinates which transform as 
$2z^2-x^2-y^2$, $x^2-y^2$, $yz$, $zx$, $xy$, respectively, under symmetric operations
(for the definition of the coordinates see Fig. \ref{Fig:C60})
and $p_\gamma$ $(\gamma = \theta, \epsilon, \xi, \eta, \zeta)$ is the conjugate momentum of $q_\gamma$.
The components $\theta, \epsilon, \xi, \eta, \zeta$ of $h_g$ mode 
are denoted 1,4,5,2,3, respectively, in Ref. \cite{S_OBrien1996}.

The $h_g$ normal coordinates can be transformed into polar coordinates $(q, \alpha, \gamma, \theta, \phi)$ \cite{S_OBrien1996}.
Under appropriate rotation of the electronic coordinates $\lambda = x,y,z$,
\begin{eqnarray}
 \hat{c}^\dagger_{l \sigma} &=& \sum_{\lambda'=x,y,z} S_{l\lambda}(\gamma, \theta, \phi) \hat{c}^\dagger_{\lambda \sigma},
\end{eqnarray}
with 
\begin{eqnarray}
 S_{l\lambda}(\gamma, \theta, \phi) &=& \left[B_P(\gamma)C_P(\theta)D_P(\phi)\right]_{l\lambda}, 
\end{eqnarray}
we obtain adiabatic electronic states $l=1,2,3$. 
Here, the rotation matrices $B_P, C_P, D_P$ are the same as in Ref. \cite{S_OBrien1996} 
(angle $\theta$ and the component $\theta$ of $h_g$ coordinate are different from each other).
By the unitary transformation 
\begin{eqnarray}
 \tilde{H}_\text{JT} = \hat{S}^\dagger \hat{H}_\text{JT} \hat{S},
\label{Eq:SHS}
\end{eqnarray}
the potential term of the JT Hamiltonian (\ref{Eq:UJT}) becomes diagonal:
\begin{eqnarray}
 U_\text{JT} &=& \frac{\hslash \omega}{2} q^2 + \hslash \omega g q
 \sum_\sigma \left[
   \cos\left(\alpha + \frac{\pi}{3}\right) \hat{n}_{1\sigma} 
   \right.
\nonumber\\
 &+&
   \left.
   \cos\left(\alpha - \frac{\pi}{3}\right) \hat{n}_{2\sigma} 
 - \cos\alpha \hat{n}_{3\sigma} 
 \right].
\label{Eq:UJT_ad}
\end{eqnarray}
The range of $\alpha$ is $0 \ge \alpha < \pi/3$ or equivalent range in the configuration space.

Minimizing Eq. (\ref{Eq:UJT_ad}) under the condition of $2(n_1 + n_2 + n_3) = n$, 
the JT deformation and the JT stabilization energy are obtained as follows \cite{S_Auerbach1994}.
Here, $n_l$ is an occupation number of electron ($n_l = 0,1$).
For example, when there is one electrons in the LUMO orbitals (C$_{60}^{-}$), 
the amplitude of the JT coordinates at the minima of the adiabatic potential energy surface (\ref{Eq:UJT_ad}) is 
\begin{eqnarray}
 (q, \alpha) &=& (g, 0),
\end{eqnarray}
with the occupation numbers
\begin{eqnarray}
 (n_1, n_2, n_3) = (0, 0, 1).
\end{eqnarray}
The JT stabilization energy (the gain by the deformation) is 
\begin{eqnarray}
 E_\text{JT} &=& E_\text{JT}^{(1)} = \frac{\hslash \omega g^2}{2}.
\label{Eq:EJT1}
\end{eqnarray}
In the case of C$_{60}$ anion, the effective $g = 1.07$ and $\omega = 87.7$ meV, 
the stabilization energy $E_\text{JT}^{(1)} = 50.2$ meV.

By the same procedure, we obtain the JT deformations, occupations, and JT stabilization energies for all cases (Table \ref{Table:JT}).

\begin{table}[tbh]
 \caption{JT distortion in polar and Cartesian coordinates, occupation numbers, and JT stabilization energies (meV) of C$_{60}^{n-}$.
 For the calculations of the Cartesian coordinates, we set the three Euler angles $\gamma, \theta, \phi$ zero,
 leading to $q_\xi=q_\eta=q_\zeta=0$.}
 \label{Table:JT}
 \begin{ruledtabular}
  \begin{tabular}{ccccc}
   $n$ & $(q,\alpha)$ & $(q_\theta, q_\epsilon)$ & $(n_1, n_2, n_3)$ & $E_\text{JT}$\\ 
   \hline
    1  & $(g,0)$             & $(g,0)$         & (0,0,1) & $E_\text{JT}^{(1)}$  \\
    2  & $(2g,0)$            & $(2g,0)$        & (0,0,2) & $4E_\text{JT}^{(1)}$ \\
    3  & $(\sqrt{3}g,\pi/2)$ & $(0,\sqrt{3}g)$ & (2,0,1) & $3E_\text{JT}^{(1)}$ \\
    4  & $(2g,\pi)$          & $(-2g,0)$       & (2,2,0) & $4E_\text{JT}^{(1)}$ \\
    5  & $(g,\pi)$           & $(-g,0)$        & (2,2,1) & $E_\text{JT}^{(1)}$  \\
  \end{tabular}
 \end{ruledtabular}
\end{table}

\begin{figure*}[tb]
\begin{center}
 \includegraphics[height=16cm, angle=-90, bb= 0 0 1444 5812]{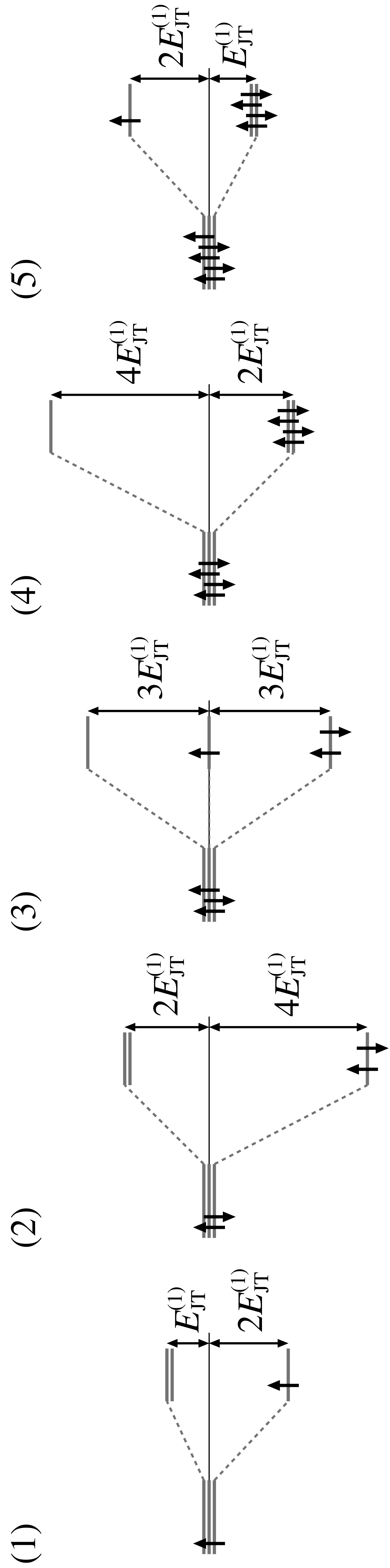}
\end{center}
\caption{JT splitting of C$_{60}^{n-}$. $E_\text{JT}^{(1)}=50.2$ meV is the JT stabilization energy of C$_{60}^{-}$.}
\label{Fig:JT}
\end{figure*}

In the strong JT coupling limit ($g\rightarrow \infty$), 
the ground state is well described in the space of the ground adiabatic state. 
Within the approximation, the kinetic term of Eq. (\ref{Eq:SHS}) can be 
separated into radial and rotational parts in the configuration space of the $h_g$ mode \cite{S_OBrien1996}.
In the case of $n=1,2,4,5$, there are three dimensional radial part and two dimensional rotational part
(Eqs. (12) and (24) in Ref. \cite{S_OBrien1996}):
for $n=1,5$,
\begin{eqnarray}
 \tilde{H}_\text{KE} &=& -\frac{\hslash \omega}{2}
 \left[
  \frac{1}{q^2}\frac{\partial}{\partial q}\left(q^2\frac{\partial}{\partial q}\right)
 + \frac{1}{q^2\sin\alpha}\frac{\partial}{\partial \alpha}\left(\sin\alpha \frac{\partial}{\partial \alpha}\right)
 \right.
\nonumber\\
 &+& 
 \left.
  \frac{1}{q^2\sin^2\alpha}\frac{\partial^2}{\partial\gamma^2} 
 \right]
\nonumber\\
 &-& \frac{\hslash\omega}{6q^2}
 \left[
  \frac{1}{\sin\theta}\frac{\partial}{\partial \theta}\left(\sin\theta \frac{\partial}{\partial \theta}\right)
 + \frac{1}{\sin^2\theta}\frac{\partial^2}{\partial\phi^2}
 \right],
\end{eqnarray}
and for $n=2,4$,
\begin{eqnarray}
 \tilde{H}_\text{KE} &=& 
 -\frac{\hslash \omega}{2}\left[
  \frac{1}{q^2}\frac{\partial}{\partial q}\left(q^2\frac{\partial}{\partial q}\right)
 + \frac{1}{q^2\sin\alpha} \frac{\partial}{\partial \alpha}\left(\sin\alpha \frac{\partial}{\partial \alpha}\right)
 \right.
\nonumber\\
 &+& 
 \left.
   \frac{1}{q^2\sin^2\alpha}\frac{\partial^2}{\partial \gamma^2}
 \right]
 + \frac{\hslash\omega}{3q^2}
\nonumber\\
 &-&
 \frac{\hslash \omega}{6q^2}
 \left[
  \frac{1}{\sin\theta}\frac{\partial}{\partial \theta}\left(\sin\theta \frac{\partial}{\partial \theta}\right)
 + \frac{1}{\sin^2\theta}\frac{\partial^2}{\partial \phi^2}
 \right].
\end{eqnarray}
Here, the coordinates $q,\alpha,\gamma$ are the radial coordinates and $\theta, \phi$ are the rotational coordinates.
Therefore, in the strong coupling limit, the ground state is described by the product of the radial and rotational wave functions,
\begin{eqnarray}
 \Psi_0 &=& \Phi_0^\text{el}(\alpha, \gamma, \theta, \phi) \phi^\text{vib}(q,\alpha,\gamma)\phi^\text{rot}(\theta, \phi),
\end{eqnarray}
and the ground eigen energy of $\tilde{H}_\text{JT}$ is 
\begin{eqnarray}
 E_0 &=& -E_\text{JT} + \frac{3}{2}\hslash \omega.
\label{Eq:E0_1245}
\end{eqnarray}
The first term of the right hand side is the stabilization by the static JT distortion,
whereas the right hand side includes the effect of the JT dynamics.
Compared with the ground energy of the five-dimensional Harmonic oscillator, $5\hslash\omega/2$, 
the JT dynamics stabilizes the system by $\hslash \omega$ because of the two rotational modes
in the minima of the adiabatic potential energy surface.

On the other hand, when $n=3$, there are two dimensional radial part and three dimensional rotational part
(Eq. (32) in Ref. \cite{S_OBrien1996}):
\begin{eqnarray}
 \tilde{H}_\text{KE} &=& -\frac{\hslash \omega}{2} \left[
  \frac{1}{q} \frac{\partial}{\partial q} \left(q\frac{\partial}{\partial q}\right)
 + \frac{1}{q^2}\frac{\partial^2}{\partial \alpha^2} + \frac{9}{4q^2}
 \right]
\nonumber\\
 &-& 
  \frac{\hslash \omega}{8q^2}\left[4\lambda_x^2+4\lambda_y^2+\lambda_z^2\right],
\end{eqnarray}
where $\lambda_x, \lambda_y, \lambda_z$ are the angular momenta described by angles $\gamma, \theta, \phi$ \cite{S_OBrien1996}.
In the ground state, the eigenstate of $H_\text{JT}$ (vibronic state) is written as the product of the
radial vibration around the minima and the 
\begin{eqnarray}
 \Psi_0 &=& \Phi_0^\text{el}(\alpha, \gamma, \theta, \phi) \phi^\text{vib}(q,\alpha)\phi^\text{rot}(\gamma, \theta, \phi),
\end{eqnarray}
and the ground eigen energy of $\tilde{H}_\text{JT}$ is 
\begin{eqnarray}
 E_0 &=& -E_\text{JT} + \hslash \omega.
\label{Eq:E0_3}
\end{eqnarray}
Compared with the zero vibrational energy of the five-dimensional Harmonic oscillator,
there is gain by $3\hslash \omega/2$ due to the JT dynamics.
In Ref. \cite{S_Iwahara2013}, the gain by the JT dynamics is evaluated ca 90 meV for C$_{60}^{3-}$. 
On the other hand, the frequency for the effective mode is 87.7 meV, and the dynamical component of the 
ground energy $3\hslash \omega/2 = 132$ meV. 
The discrepancy is due to the intermediate strength of the orbital vibronic coupling constant $g$ of C$_{60}^{n-}$ anion.
In the main text, we assume that, however, the relative strength of the dynamical JT stabilization is the same, 
and estimate the gain of C$_{60}^{4-}$.

\section{Gutzwiller's approach to Jahn-Teller systems}
\subsection{Self-consistent Gutzwiller wave function}
We briefly explain the Gutzwiller's approach to static Jahn-Teller system developed in Ref. \cite{S_Iwahara2015}.
The Gutzwiller's wave function $\Phi_\text{G}$ is written as 
\begin{eqnarray}
 |\Phi_\text{G}\rangle &=& \hat{P}_\text{G}|\Phi_\text{S}\rangle,
\label{Eq:PhiG}
\end{eqnarray}
where, $\Phi_\text{S}$ is a Slater determinant,
\begin{eqnarray}
 |\Phi_\text{S}\rangle &=& \prod_{\alpha \mathbf{k} \sigma}^{\text{occ.}} \hat{a}_{\alpha \mathbf{k} \sigma}^\dagger |0\rangle,
\label{Eq:PhiS}
\end{eqnarray}
and $\hat{P}_\text{G}$ is Gutzwiller's projector,
\begin{eqnarray}
 \hat{P}_\text{G} &=& \exp\left[
 -\frac{1}{2}\sum_\mathbf{m} \sum_{\lambda \sigma \ne \lambda' \sigma'} A_{\lambda \lambda'} 
 \hat{n}_{\lambda \mathbf{m}\sigma}
 \hat{n}_{\lambda' \mathbf{m}\sigma'}
 \right].
\label{Eq:PG}
\end{eqnarray}
Here, we assume the translational symmetry of the system, and the state $\alpha \mathbf{k}$ 
is a linear combination of the localized states $\lambda \mathbf{m}$:
\begin{eqnarray}
 \hat{a}_{\alpha \mathbf{k} \sigma}^\dagger &=&
 \sum_\mathbf{m} \frac{e^{i\mathbf{k}\cdot \mathbf{m}}}{\sqrt{N}} 
 u_{\lambda, \alpha \mathbf{k}}
 \hat{c}_{\lambda \mathbf{m} \sigma}^\dagger,
\label{Eq:a+}
\end{eqnarray}
with orbital coefficients $u_{\lambda, \alpha \mathbf{k}}$.
$N$ is the number of sites in the system.
In Eq. (\ref{Eq:PG}), $A_{\lambda \lambda'}$ are real variational parameters.
The Gutzwiller's variational parameter $A_{\lambda \lambda'}$ are orbital specific 
in order to adequately treat the split orbitals.

With the use of the Gutzwiller's wave function (\ref{Eq:PhiG}),
the ground state energy per site $E_g$ was calculated. 
The latter consists of the band energy $E_\text{band}$, 
linear Jahn-Teller energy $U_\text{JT}$, elastic energy $U_\text{el}$, and bielectronic energy $E_\text{bi}$:
\begin{eqnarray}
 E_g &=& E_\text{band} + E_\text{bi} + E_\text{JT}.
\end{eqnarray}
The band energy is written as 
\begin{eqnarray}
 E_\text{band} &=& \sum_{\lambda \lambda' \sigma} q_{\lambda \lambda'} \tau_{\lambda \lambda'},
\end{eqnarray}
where $q_{\lambda \lambda'}$ is Gutzwiller's reduction factor, 
which has the meaning of quasi-particle weight \cite{S_Gutzwiller1965}, 
and $\tau_{\lambda \lambda'}$ is the $\lambda,\lambda'$ element of the uncorrelated band energy.
For the calculation of $E_\text{band}$ we used Gutzwiller's approximation \cite{S_Gutzwiller1963, S_Ogawa1975}.
The Jahn-Teller energy for C$_{60}^{3-}$ is 
\begin{eqnarray}
 E_\text{JT} &=& \frac{\hslash \omega}{2}q^2 - \sum_\sigma \frac{\sqrt{3}}{2}\hslash \omega g q(n_x - n_y),
\label{Eq:UJT_3}
\end{eqnarray}
and for C$_{60}^{4-}$,
\begin{eqnarray}
 E_\text{JT} &=& \frac{\hslash \omega}{2}q^2 - \sum_\sigma \hslash \omega g q(-2n_x + n_y + n_z).
\label{Eq:UJT_4}
\end{eqnarray}
Here, we assume that the Jahn-Teller distortion (Table \ref{Table:JT}) is common to all of the fullerene sites,
and $q$ is the magnitude of the deformation.
For the bielectronic energy, see Ref. \cite{S_Iwahara2015}.

The total energy $E_g$ contains two types of the variational parameters:
orbital coefficients $u$ (\ref{Eq:a+}) and Gutzwiller's parameter $A$ (\ref{Eq:PG}). 
The energy $E_g$ is minimized with respect to both ${u}$ and ${A}$. 
Variational calculations of the energy are performed separately for ${u}$ an ${A}$. 
From the variation of the energy with respect to ${u}$ with fixed ${A}$, 
we obtain Hartree-Fock like equation for each $\mathbf{k}$:
\begin{eqnarray}
 \sum_{\lambda'} h^\mathbf{k}_{\lambda \lambda'} u_{\lambda',\alpha \mathbf{k}} 
 &=& \epsilon_{\alpha \mathbf{k}} u_{\lambda,\alpha \mathbf{k}},
\label{Eq:hk}
\end{eqnarray}
where $h^\mathbf{k}_{\lambda \lambda'}$ is one-electron Hamiltonian, 
$\epsilon_{\alpha \mathbf{k}}$ is one-electron eigen energy of the Hamiltonian.
On the other hand, for fixed ${u}$, we minimize $E_g$ with respect to ${A}$:
\begin{eqnarray}
 \frac{\partial E_g}{\partial A_{\lambda \lambda'}} &=& 0.
\label{Eq:dEGdA}
\end{eqnarray}
These two equations (\ref{Eq:hk}), (\ref{Eq:dEGdA}) are solved repeatedly until we obtain the convergence of the energy. 
During the self-consistent calculation, the populations $n_\lambda$ are fixed,
and the ground state for each set of $\{n_\lambda\}$ was performed.
For details of the self-consistent Gutzwiller's approach, see Ref. \cite{S_Iwahara2015}.

\subsection{Ground state of fcc K$_3$C$_{60}$}
\begin{figure}[htb]
 \begin{center}
   \includegraphics[height=7cm, bb= 0 0 3306 2861]{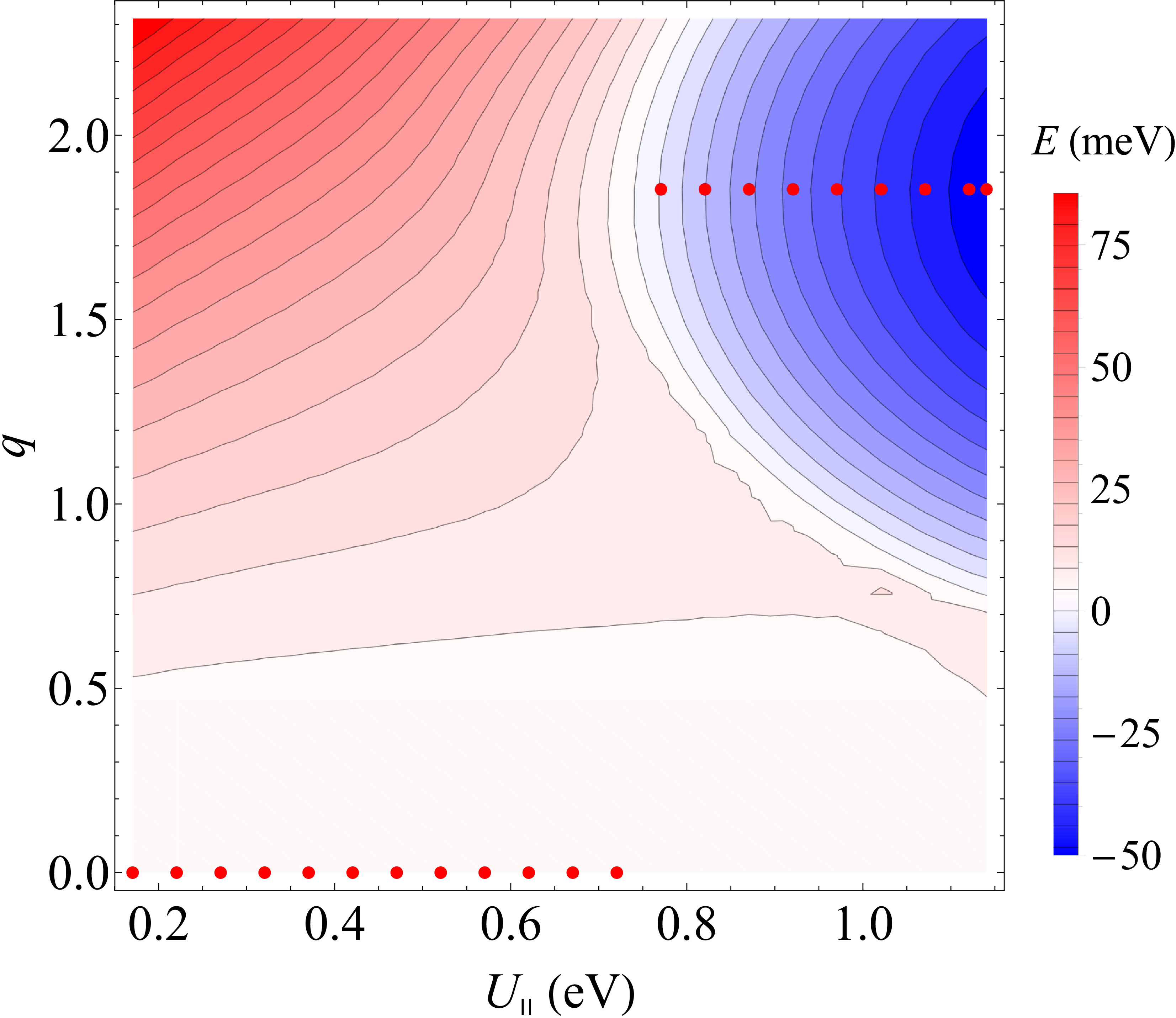}
 \end{center}
 \caption{Ground state energy of K$_3$C$_{60}$.
  The total energy is plotted as a function of the Jahn-Teller deformation $q$ and $U_\parallel$ (in meV).
  For each $U_\parallel$, the energy $E_g$ at $q=0$ is subtracted from $E_g$.
  The red points show the ground state for each $U_\parallel$. 
  }
\label{Fig:EqU_K3C60}
\end{figure}

The ground state energy as a function of the Coulomb repulsion energy $U_\parallel$ 
and the Jahn-Teller distortion $q$ is shown in Fig. \ref{Fig:EqU_K3C60}.
In the figure, the ground state for each $U_\parallel$ is indicated by red point. 
For small $U_\parallel$, the JT distortion is suppressed, 
whereas for $U_\parallel \agt 0.75$ eV, the JT distortion is favored.
Therefore, in the former region of $U_\parallel$, the orbitals are degenerate and equally populated in the ground state, 
while in the latter region, the LUMO levels are completely split and orbital disproportionation arises.

\section{Tight binding model of bct K$_4$C$_{60}$}
\subsection{Tight binding parametrization}
\begin{figure}[htb]
\begin{center}
\includegraphics[height=5cm, bb= 0 0 3517 2511]{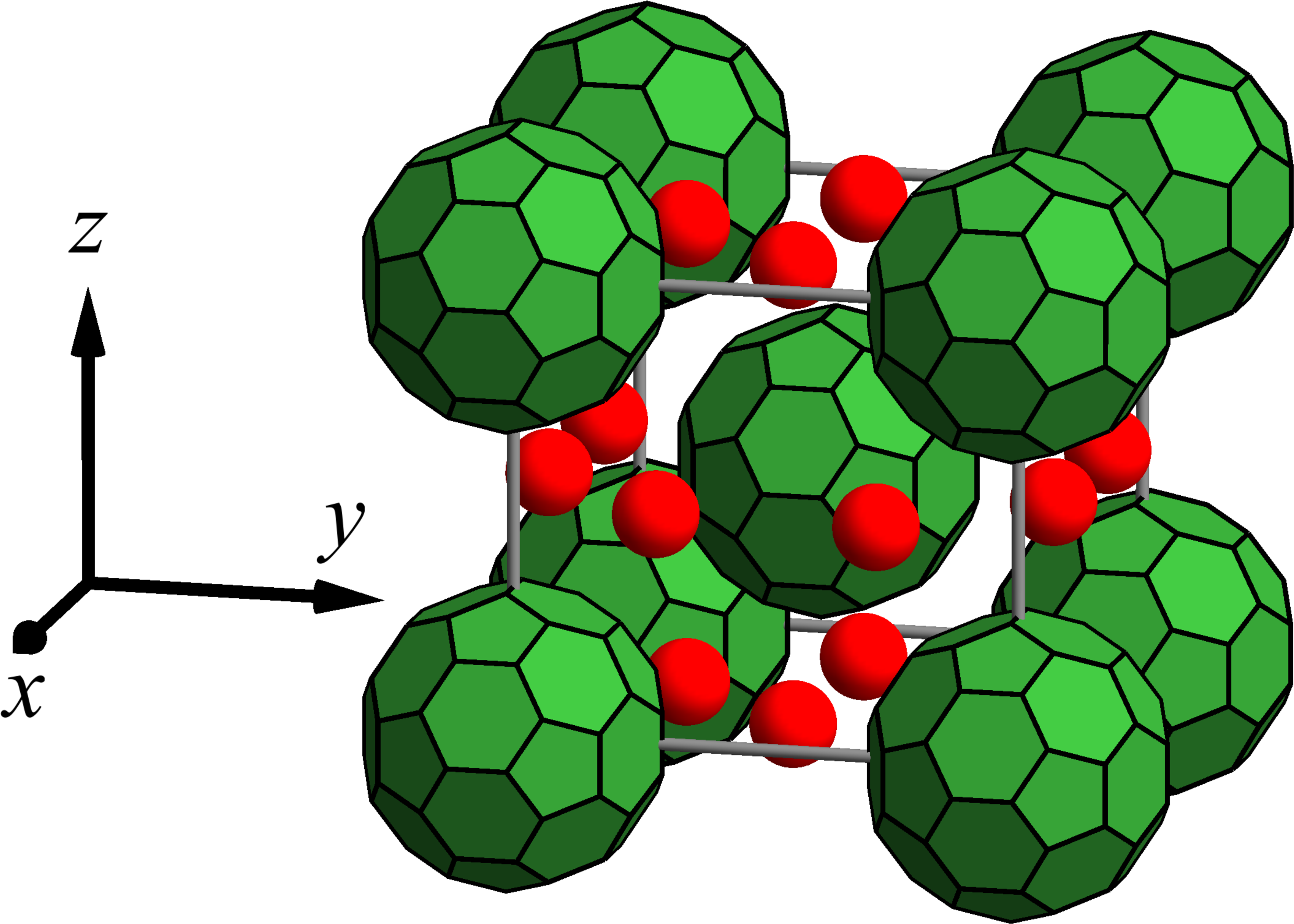}
\end{center}
\caption{
\label{Fig:bct_K4C60}
Structure of bct K$_4$C$_{60}$. 
The green balls are fullerene C$_{60}$ and the red spheres are K atoms which locate at $a(1/2,0.218152,0)$ 
and equivalent positions with respect to the central C$_{60}$.
}
\end{figure}

K$_4$C$_{60}$ has body centered tetragonal (bct) structure (Fig. \ref{Fig:bct_K4C60}).
The primitive lattice vector is 
\begin{eqnarray}
 \mathbf{a}_1 &=& \frac{1}{2}(a,-a,c), \quad
 \mathbf{a}_2 = \frac{1}{2}(a,a,c), 
\nonumber\\
 \mathbf{a}_3 = \frac{1}{2}(-a,-a,c),
\label{Eq:a}
\end{eqnarray}
where, $a$ and $c$ are the lattice constants.
In the present work, the lattice constants were taken from the neutron diffraction data measured at 6 K 
($a = 11.827$ $\text{\AA}$, $c = 10.746$ $\text{\AA}$) \cite{S_Klupp2006}.
Using $\mathbf{a}_1, \mathbf{a}_2, \mathbf{a}_3$, the nearest neighbor sites (Fig. \ref{Fig:bct_K4C60}) are described as 
\begin{eqnarray}
 \Delta \mathbf{m} &=& 
  \pm \mathbf{a}_1, \pm \mathbf{a}_2, \pm (-\mathbf{a}_1+\mathbf{a}_2+\mathbf{a}_3), \pm \mathbf{a}_3.
\label{Eq:nn}
\end{eqnarray}
The next nearest neighbor sites are  
\begin{eqnarray}
 \Delta \mathbf{m} &=& 
 \pm a \mathbf{e}_x, 
 \pm a \mathbf{e}_y, 
 \pm c \mathbf{e}_z, 
\end{eqnarray}
where $\mathbf{e}_x, \mathbf{e}_y, \mathbf{e}_z$ are the unit vectors along the axes $x,y,z$, respectively.

Because of the lower symmetry of bct lattice than fcc one, the orbital energy levels of each site split into three. 
Thus, the tight-binding model Hamiltonian for the bct lattice is written as 
the sum of the orbital energy levels part, 
the electron transfer part between the nearest-neighbor (nn) sites and that between next nearest neighbor (nnn) sites:
\begin{eqnarray}
 \hat{H}_\text{t} &=& \sum_\mathbf{m} \sum_{\lambda \sigma} \epsilon_\lambda \hat{n}_{\lambda \mathbf{m} \sigma} + 
                      \sum_\mathbf{m} \sum_\sigma \left(\hat{H}_{\mathbf{m}\sigma}^\text{nn} 
                      + \hat{H}_{\mathbf{m}\sigma}^\text{nnn}\right), 
\label{Eq:Ht_bct}
\end{eqnarray}
where the nearest neighbor term is 
\begin{widetext}
\begin{eqnarray}
 \hat{H}_{\mathbf{m}\sigma}^\text{nn} &=&
  \sum_{i=1}^4 
 \left(
  t_{xx}
  \hat{c}_{\mathbf{m}+\Delta \mathbf{m}_i x \sigma}^\dagger \hat{c}_{x\mathbf{m} \sigma} 
 + t_{yy}
  \hat{c}_{\mathbf{m}+\Delta \mathbf{m}_i y \sigma}^\dagger \hat{c}_{y\mathbf{m} \sigma} 
+
 t_{zz}
  \hat{c}_{\mathbf{m}+\Delta \mathbf{m}_i z \sigma}^\dagger \hat{c}_{z\mathbf{m} \sigma} 
 \right)
\nonumber\\
 &+& \sum_{(\lambda,\lambda') = (y,x), (x,y)} t_{xy}
 \left(
  - \hat{c}_{\lambda \mathbf{m}+\Delta \mathbf{m}_1 \sigma}^\dagger \hat{c}_{\lambda' \mathbf{m} \sigma} 
  + \hat{c}_{\lambda \mathbf{m}+\Delta \mathbf{m}_2 \sigma}^\dagger \hat{c}_{\lambda' \mathbf{m} \sigma} 
-
    \hat{c}_{\lambda \mathbf{m}+\Delta \mathbf{m}_3 \sigma}^\dagger \hat{c}_{\lambda' \mathbf{m} \sigma} 
  + \hat{c}_{\lambda \mathbf{m}+\Delta \mathbf{m}_4 \sigma}^\dagger \hat{c}_{\lambda' \mathbf{m} \sigma} 
 \right)
\nonumber\\
 &+& \sum_{(\lambda,\lambda') = (y,z), (z,y)} t_{yz}
 \left(
  - \hat{c}_{\lambda\mathbf{m}+\Delta \mathbf{m}_1 \sigma}^\dagger \hat{c}_{\lambda'\mathbf{m} \sigma} 
  + \hat{c}_{\lambda\mathbf{m}+\Delta \mathbf{m}_2 \sigma}^\dagger \hat{c}_{\lambda'\mathbf{m} \sigma} 
+
  + \hat{c}_{\lambda \mathbf{m}+\Delta \mathbf{m}_3 \sigma}^\dagger \hat{c}_{\lambda' \mathbf{m} \sigma} 
  - \hat{c}_{\lambda \mathbf{m}+\Delta \mathbf{m}_4 \sigma}^\dagger \hat{c}_{\lambda' \mathbf{m} \sigma} 
 \right)
\nonumber\\
 &+& \sum_{(\lambda,\lambda') = (z,x), (x,z)} t_{zx}
 \left(
    \hat{c}_{\lambda \mathbf{m}+\Delta \mathbf{m}_1 \sigma}^\dagger \hat{c}_{\lambda' \mathbf{m} \sigma} 
  + \hat{c}_{\lambda \mathbf{m}+\Delta \mathbf{m}_2 \sigma}^\dagger \hat{c}_{\lambda' \mathbf{m} \sigma} 
-
    \hat{c}_{\lambda \mathbf{m}+\Delta \mathbf{m}_3 \sigma}^\dagger \hat{c}_{\lambda' \mathbf{m} \sigma} 
  - \hat{c}_{\lambda \mathbf{m}+\Delta \mathbf{m}_4 \sigma}^\dagger \hat{c}_{\lambda' \mathbf{m} \sigma} 
 \right)
\nonumber\\
 &+& \text{H.c.}
\label{Eq:Htnn}
\end{eqnarray}
and the next nearest neighbor term is 
\begin{eqnarray}
 \hat{H}_{\mathbf{m}\sigma}^\text{nnn} &=& 
     t_{xx}^{\prime x} \hat{c}_{x\mathbf{m}+\mathbf{e}_x\sigma}^\dagger \hat{c}_{x\mathbf{m}\sigma} 
  -  t_{yy}^{\prime x} \hat{c}_{y\mathbf{m}+\mathbf{e}_x\sigma}^\dagger \hat{c}_{y\mathbf{m}\sigma} 
 - t_{zz}^{\prime x} \hat{c}_{z\mathbf{m}+\mathbf{e}_x\sigma}^\dagger \hat{c}_{z\mathbf{m}\sigma} 
\nonumber\\
 &-& t_{xx}^{\prime y} \hat{c}_{x\mathbf{m}+\mathbf{e}_y\sigma}^\dagger \hat{c}_{x\mathbf{m}\sigma} 
  +  t_{yy}^{\prime y} \hat{c}_{y\mathbf{m}+\mathbf{e}_y\sigma}^\dagger \hat{c}_{y\mathbf{m}\sigma} 
 - t_{zz}^{\prime y} \hat{c}_{z\mathbf{m}+\mathbf{e}_y\sigma}^\dagger \hat{c}_{z\mathbf{m}\sigma} 
\nonumber\\
 &-& t_{xx}^{\prime z} \hat{c}_{x\mathbf{m}+\mathbf{e}_z\sigma}^\dagger \hat{c}_{x\mathbf{m}\sigma} 
  -  t_{yy}^{\prime z} \hat{c}_{y\mathbf{m}+\mathbf{e}_z\sigma}^\dagger \hat{c}_{y\mathbf{m}\sigma} 
 + t_{zz}^{\prime z} \hat{c}_{z\mathbf{m}+\mathbf{e}_z\sigma}^\dagger \hat{c}_{z\mathbf{m}\sigma} 
\nonumber\\
 &+& \text{H.c.}
\label{Eq:Htnnn}
\end{eqnarray}
\end{widetext}
Here, $\epsilon_\lambda$ is the orbital energy level, 
$t_{\lambda \lambda'}$ and $t'_{\lambda \lambda'}$ are the electron transfer parameters 
between the nearest neighbors and next nearest neighbors.

\begin{figure}[htb]
\begin{center}
\includegraphics[height=7cm, bb= 0 0 2631 2769]{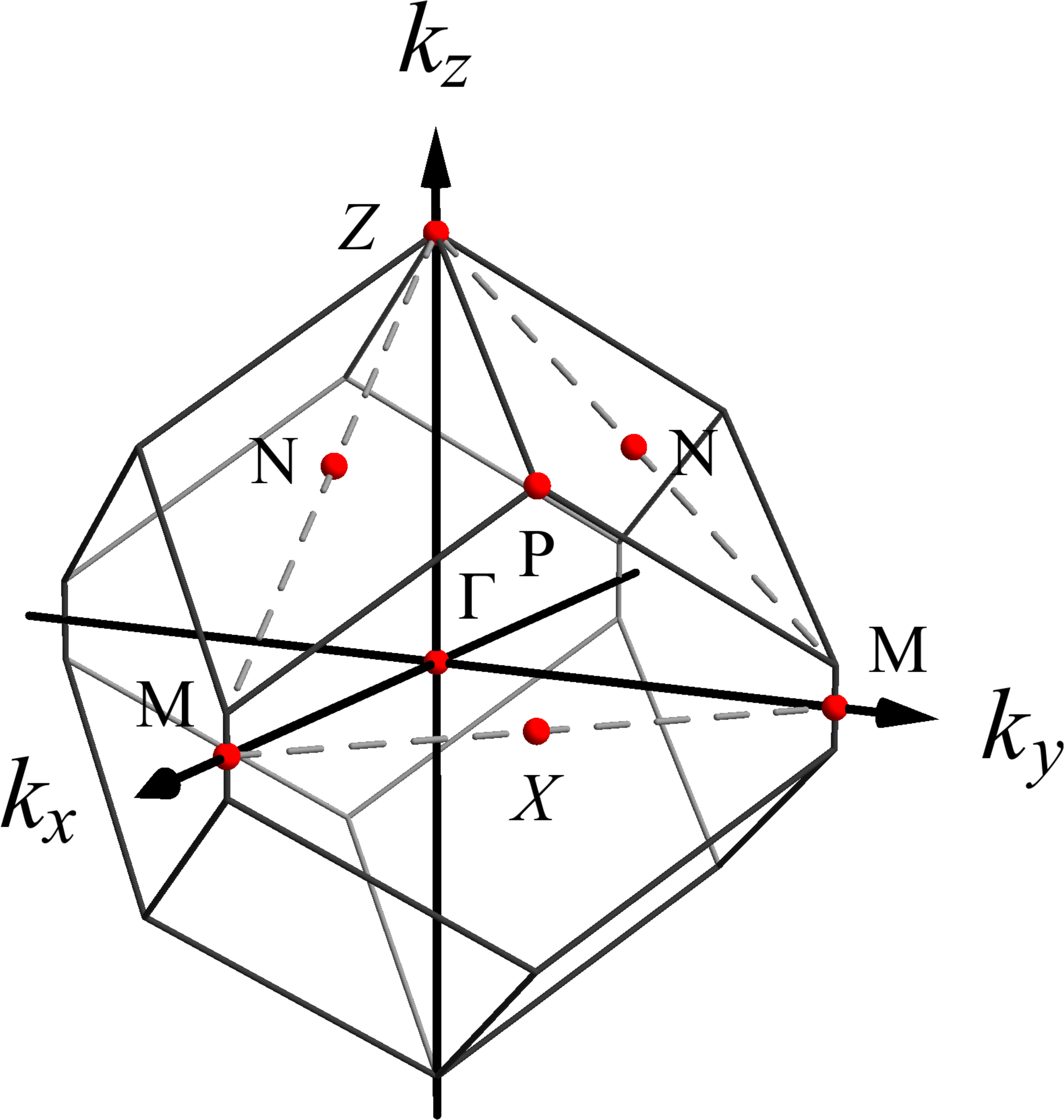}
\end{center}
\caption{
\label{Fig:IBZ}
The first Brillouin zone of bct lattice. 
The band is plotted along the path 
$Z = (0,0,(1+(a/c)^2)c/(2a))$
$\rightarrow$
$\Gamma = (0,0,0)$
$\rightarrow$
$M = (1,0,0)$
$\rightarrow$
$P = (1/2,1/2,c/(2a))$
$\rightarrow$
$M = (0,1,0)$
$\rightarrow$
$\Gamma$
$\rightarrow$
$X = (1/2,1/2,0)$
$\rightarrow$
$P$
$\rightarrow$
$N=(1/2,0,c/(2a))$
$\rightarrow$
$Z$
$\rightarrow$
$N=(0,1/2,c/(2a))$
$\rightarrow$
$P$
$\rightarrow$
$Z$
in the unit of $2\pi/a$.
}
\end{figure}

We obtained the orbital energy levels $\epsilon$ and transfer parameters $t$ 
from the fitting of the DFT band structure to the tight-binding model Hamiltonian
(\ref{Eq:Ht_bct}), (\ref{Eq:Htnn}), and (\ref{Eq:Htnnn}).
The transfer parameters $t_{\lambda \lambda'}^{\Delta \mathbf{m}}$ were taken from 
Ref. \cite{S_Iwahara2015} for fcc K$_3$C$_{60}$ and derived from the DFT calculations for bct K$_4$C$_{60}$.
The DFT calculations were peformed within the generalized gradient approximation (GGA) 
with the pseudopotentials C.pbe-mt\_fhi.UPF and K.pbe-mt\_fhi.UPF of QUANTUM ESPRESSO 5.1 \cite{S_QE-2009}.
The nuclear positions were relaxed, whereas the lattice constants from Ref. \cite{S_Klupp2006} were fixed.
The tight-binding parameters were obtained by fitting the DFT band to the model transfer Hamiltonian ($\hat{H}_\text{t}$)
incluing the nearest neighbour and next nearest neighbour terms. 
The results are shown in Fig. \ref{Fig4}(A).
For the symmetric points indicated in Fig. 4(A), see Fig. \ref{Fig:IBZ}.
Table \ref{Table:t} shows the derived parameters.
The $y$ orbital energy level is lower than the quasidegenerate $x$ and $z$ orbital levels by about 130 meV. 
We also note that the transfer parameters to the next nearest neighbor sites $t'$ are comparable to
the those of the nearest neighbor $t$. 
Particularly, the transfer parameter along the $z$ direction $t^{\prime z}_{xx}$ is the largest than the others.
This is explained by the smaller distance between C$_{60}$ sites than the other directions ($c<a$).
Therefore, the electron transfer parameters to the next nearest neighbor is crucial to describe the 
band structure of the bct A$_4$C$_{60}$.

\begin{table}[tb]
\caption{
\label{Table:t}
LUMO levels (eV) and transfer parameters (meV) of bct K$_4$C$_{60}$.
}
\begin{ruledtabular}
\begin{tabular}{ccccccccc}
$\epsilon_x$ & $\epsilon_y$ & $\epsilon_z$ & $t_{xx}$ & $t_{yy}$ & $t_{zz}$ & $t_{xy}$ & $t_{yz}$ & $t_{zx}$ \\
\hline
4.849 & 4.715 & 4.847 & 13.4 & 32.1 & 17.0 & $-$17.1 & 14.6 & 0.0 \\
\hline
$t_{xx}^{\prime x}$ & $t_{yy}^{\prime x}$ & $t_{zz}^{\prime x}$ &
$t_{xx}^{\prime y}$ & $t_{yy}^{\prime y}$ & $t_{zz}^{\prime y}$ &
$t_{xx}^{\prime z}$ & $t_{yy}^{\prime z}$ & $t_{zz}^{\prime z}$ \\
\hline
14.4 & 7.5 & $-$9.0 & $-$2.6 & 6.1 & 14.8 & 51.3 & 8.8 & 19.8 \\
\end{tabular}
\end{ruledtabular}
\end{table}

\subsection{Effect of the hybridization of LUMO bands in bct K$_4$C$_{60}$}
\begin{figure}[htb]
 \begin{center}
  \begin{tabular}{l}
   (A) \\
   \includegraphics[width=7cm, bb= 0 0 3965 2424]{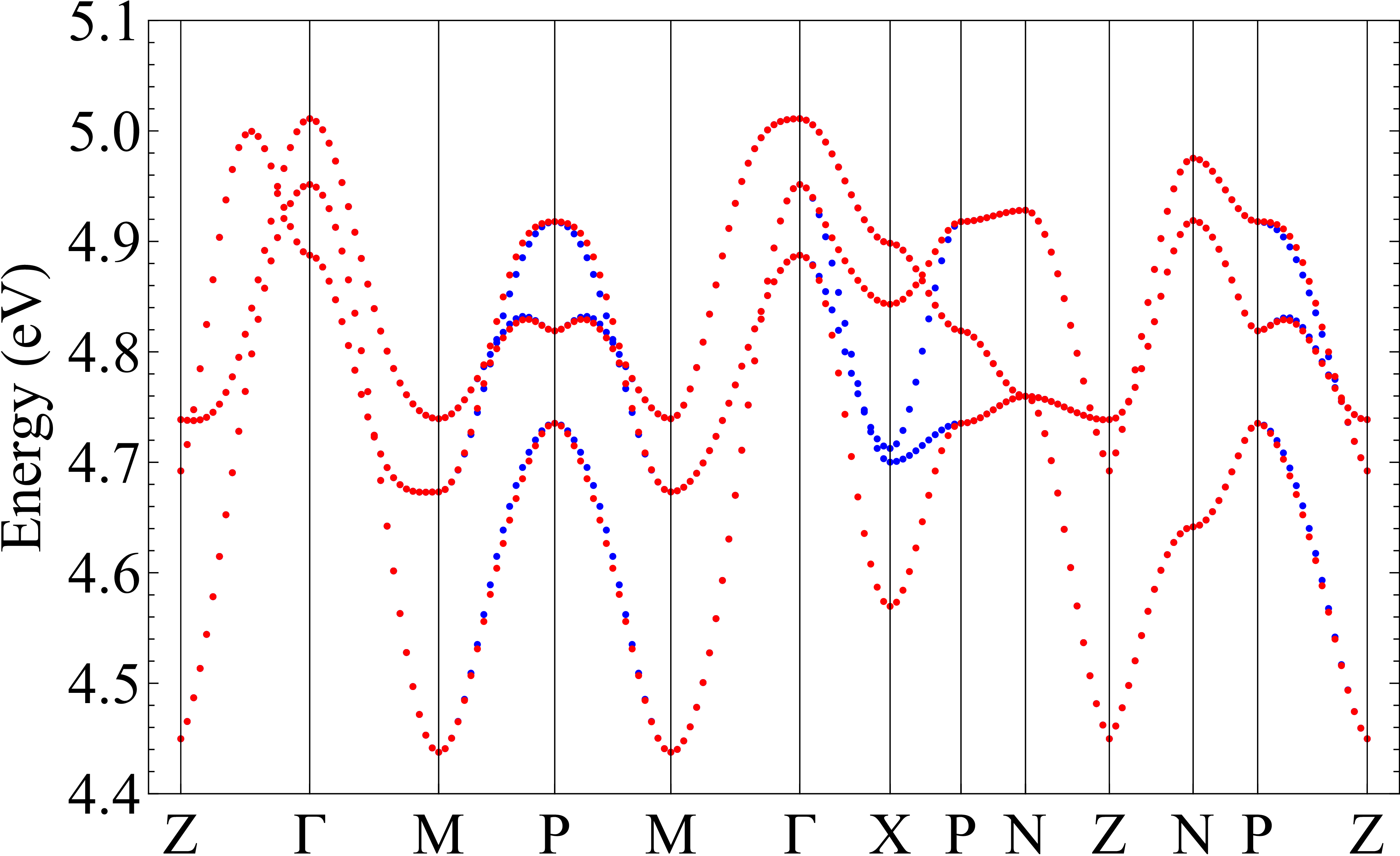}
\\
   (B) \\
   \includegraphics[width=7cm, bb= 0 0 3958 2625]{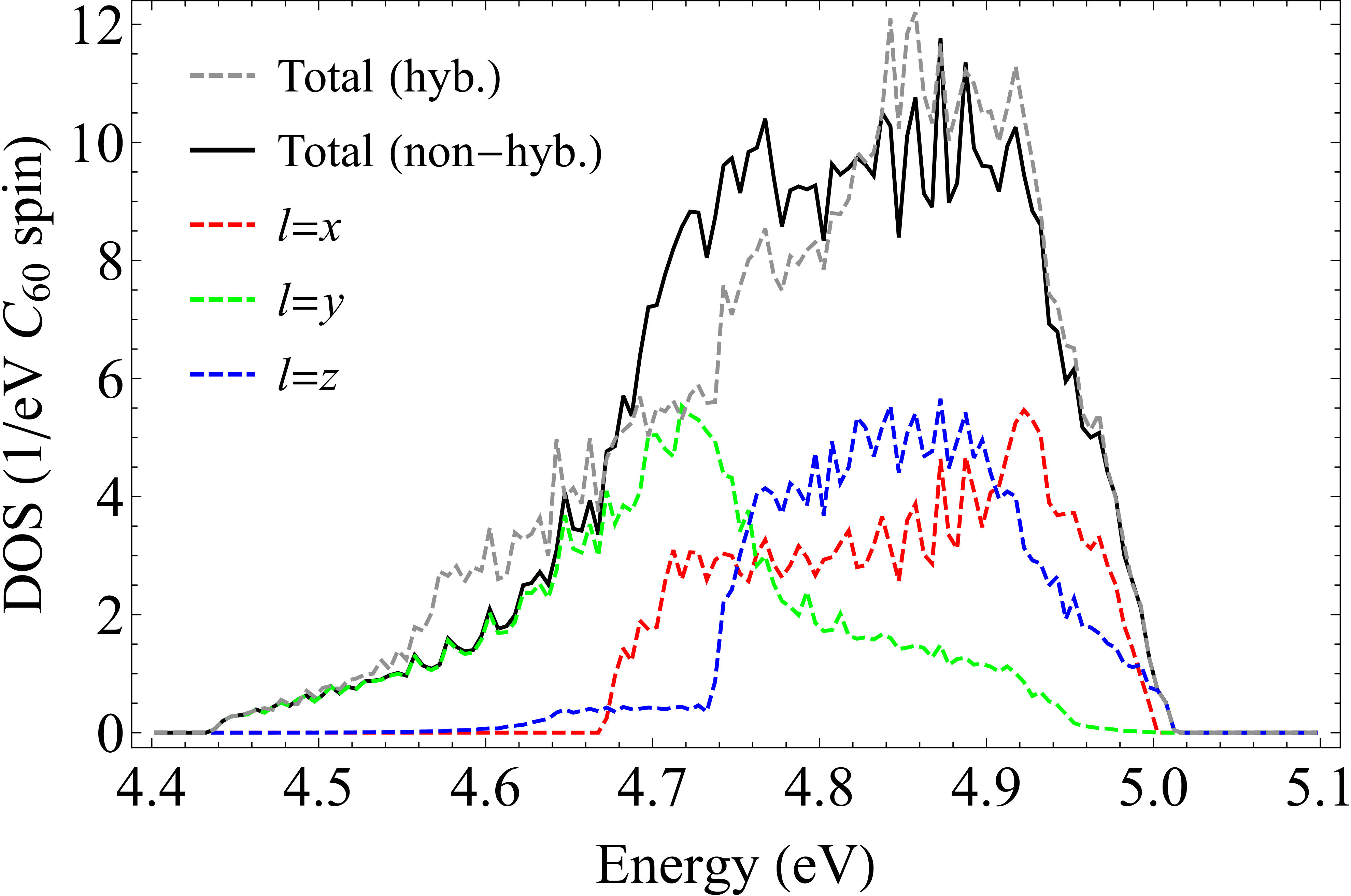}
  \end{tabular}
 \end{center}
 \caption{(A) The band structures and (B) the density of states (DOS) of bct K$_4$C$_{60}$ 
 with and without hybridization. The Jahn-Teller splitting is not taken into account. 
 (A) The red and blue dots indicate the presence and the absence of the hybridization 
 between the x orbital and the other orbitals, respectively. 
 (B) The gray dashed line is the total DOS with hybridization,
 the black solid line is the total DOS without hybridization, 
 the red, green, and blue dashed lines indicate the partial DOS. 
 }
 \label{Fig:band}
\end{figure}

\begin{figure}[htb]
 \begin{center}
   \includegraphics[width=7cm, bb= 0 0 3965 2424]{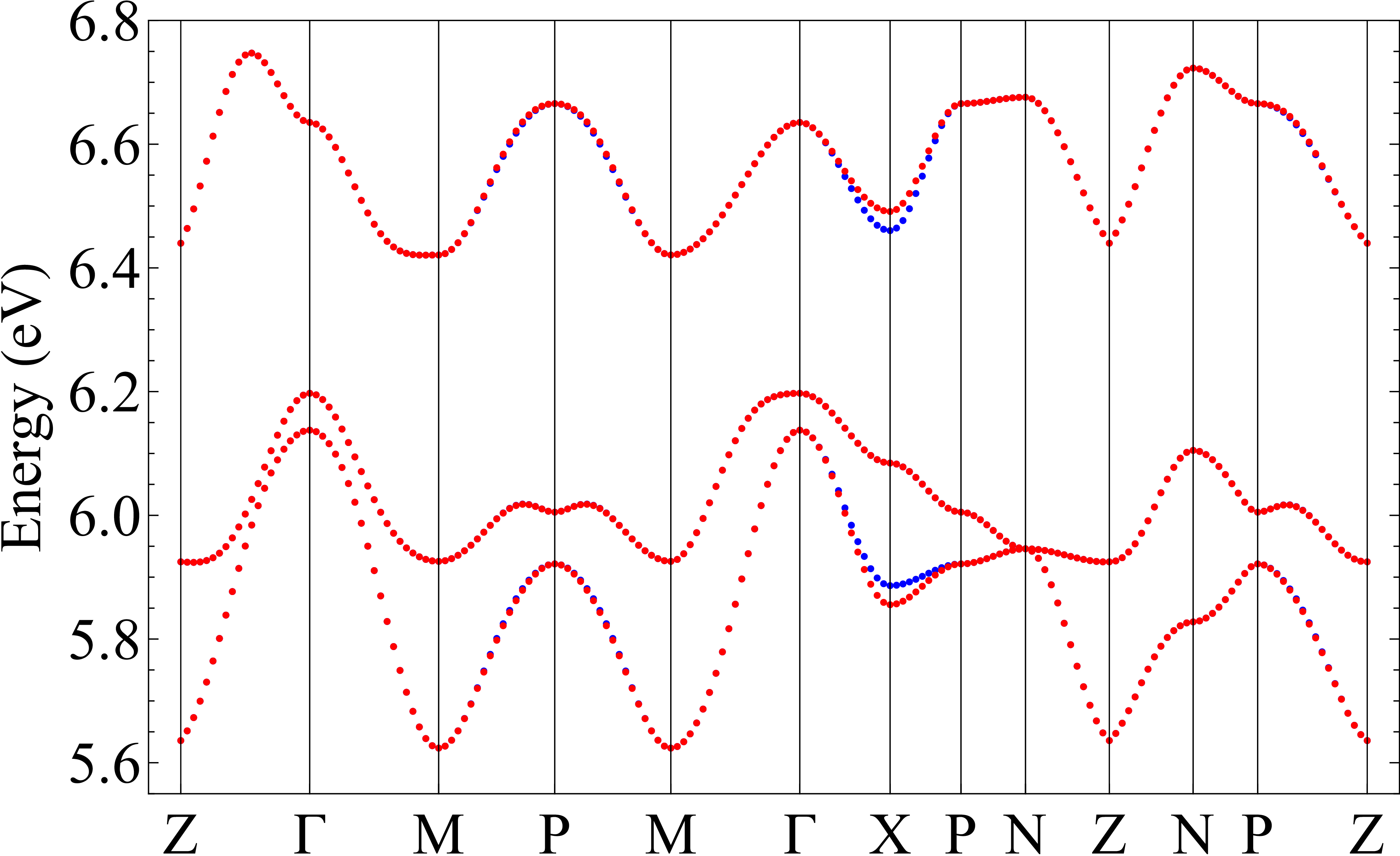}
 \end{center}
 \caption{
 The band structure with the splitting of the orbital levels.
 The hybridized (red) and non-hybridized (blue) band structures are similar to each other.
 }
 \label{Fig:split_band}
\end{figure}

In bct K$_4$C$_{60}$, the effect of the interorbital hybridization is not strong.
This is directly observed replacing the interorbital transfer parameter with zero. 
As an example, we replace the one between the $x$ orbital and the $y,z$ orbitals, ($t_{xy}, t_{zx}$).
Figure \ref{Fig:band}(A) shows the hybridized (red) and non-hybridized (blue) band structures. 
One finds that these two bands are close to each other in almost all $\mathbf{k}$-points and 
except for around the $X$ point (Fig. \ref{Fig:IBZ}).
By neglecting the hybridization, the split levels of the hybridized band (around 4.6 eV and 4.9 eV) 
become quasi-degenerate (around 4.7-4.8 eV). 
Consequently, the total density of states (DOS) for the non-hybridized band 
is enhanced around the range of 4.7-4.8 eV and is reduced around 4.6 eV and 4.9 eV (Fig. \ref{Fig:band}(B)). 

This hybridization effect is diminished by the splitting of the band 
due to the Jahn-Teller effect and Coulomb repulsion (Fig. \ref{Fig:split_band}). 
Here, we consider the Jahn-Teller distortion which would give the largest gain of the energy,
i.e., $x$ is unstabilized and $y$ and $z$ are stabilized,
and $U_\parallel = 0.5$ eV.
Therefore, bct K$_4$C$_{60}$ can be treated as a non-hybridized band system in a good approximation.

\section{Exact solution for orbitally disproportionated state and the one-particle
exciations in A$_4$C$_{60}$ with non-hybridized LUMO bands}
We show the band state and electron- and hole-quasiparticle states of 
non-hybridized multiband Hubbard Hamiltonian:
\begin{eqnarray}
 \hat{H} &=&
 \sum_{\lambda \mathbf{k} \sigma} \epsilon_{\lambda \mathbf{k}} \hat{n}_{\lambda \mathbf{k} \sigma}
 + 
 \sum_{\mathbf{m}} 
 \left[
  \sum_{\lambda} U_\parallel \hat{n}_{\lambda \mathbf{m} \uparrow} \hat{n}_{\lambda \mathbf{m} \downarrow}
 \right.
\nonumber\\
  &+&
 \left.
  (U_\perp - J) 
   \left(\hat{n}_{1\mathbf{m} \uparrow} \hat{n}_{2\mathbf{m} \uparrow} 
  + \hat{n}_{1\mathbf{m} \downarrow} \hat{n}_{2\mathbf{m} \downarrow}\right)
  \right.
\nonumber\\
 &+&
  U_\perp 
   \left(\hat{n}_{1\mathbf{m} \uparrow} \hat{n}_{2\mathbf{m} \downarrow} 
  + \hat{n}_{1\mathbf{m} \downarrow} \hat{n}_{2\mathbf{m} \uparrow}\right)
\nonumber\\
 &+&
   \sum_{\lambda \ne \lambda'} J
   \left(
    \hat{c}_{\lambda \mathbf{m}\uparrow}^\dagger
    \hat{c}_{\lambda \mathbf{m}\downarrow}^\dagger
    \hat{c}_{\lambda'\mathbf{m}\downarrow}
    \hat{c}_{\lambda'\mathbf{m}\uparrow}
  \right.
\nonumber\\
&+&
  \left.
  \left.
    \hat{c}_{\lambda \mathbf{m}\uparrow}^\dagger
    \hat{c}_{\lambda'\mathbf{m}\downarrow}^\dagger
    \hat{c}_{\lambda \mathbf{m}\downarrow}
    \hat{c}_{\lambda'\mathbf{m}\uparrow}
   \right)
 \right],
\label{Eq:Hnonhyb}
\end{eqnarray}
where, $\epsilon_{\lambda \mathbf{k}}$ is the band energy.

For simplicity, we first consider that each site has two orbitals $\lambda = 1,2$,
then consider the case with three orbitals $\lambda = 1, 2, 3$.

We assume that the ground state of the system is band insulator type, and 
orbitals $\lambda = 1$ are doubly occupied and orbitals $\lambda=2$ are empty for all $\mathbf{k}$.
The ground state of the Hamiltonian is given by 
\begin{eqnarray}
 |\Phi_0\rangle &=& \prod_{\mathbf{k}\sigma} \hat{a}_{1\mathbf{k}\sigma}^\dagger|0\rangle
\\
                    &=& \prod_{\mathbf{m}\sigma} \hat{c}_{1\mathbf{m}\sigma}^\dagger|0\rangle,
\label{Eq:Phin}
\end{eqnarray}
where, $n$ is the number of electrons in the system. 
The ground energy $E_0$ is directly calculated:
\begin{eqnarray}
 \hat{H}|\Phi_0\rangle &=& 
 \left(\sum_\mathbf{k} 2\epsilon_{1\mathbf{k}} 
  + \sum_\mathbf{m} U_{\parallel}\right) |\Phi_0\rangle,
\end{eqnarray}
therefore, 
\begin{eqnarray}
 E_0 &=& \sum_\mathbf{k} 2\epsilon_{1\mathbf{k}} + N U_{\parallel}.
\end{eqnarray}

Now, we add one electron to empty band orbital $2\mathbf{k}$:
\begin{eqnarray}
 |\Phi^e_{2\mathbf{k}\sigma}\rangle &=& \hat{a}_{2\mathbf{k}\sigma}^\dagger |\Phi_0\rangle,
\label{Eq:Phin+1}
\end{eqnarray}
This state is also an eigenstate of $\hat{H}$ (\ref{Eq:Hnonhyb}):
\begin{eqnarray}
 \hat{H}|\Phi^e_{2\mathbf{k}\sigma}\rangle &=& E_{2\mathbf{k}}|\Phi^e_{2\mathbf{k}\sigma}\rangle.
\end{eqnarray}
Moreover, as in the previous case, Eq. (\ref{Eq:Phin+1}) is an eigenstate of 
each term of the Hamiltonian (electron transfer and bielectronic parts).
The first part is obtained as 
\begin{eqnarray}
 \hat{H}_\text{t}
 |\Phi^e_{2\mathbf{k}\sigma}\rangle 
 &=&
 \sum_{\lambda' \mathbf{k}' \sigma'} \epsilon_{\lambda'\mathbf{k}' } \hat{n}_{\lambda' \mathbf{k}' \sigma'} 
 \hat{a}_{2\mathbf{k}\sigma}^\dagger|\Phi_0\rangle
\nonumber\\
 &=&
 \hat{a}_{2\mathbf{k}\sigma}^\dagger
 \sum_{\mathbf{k}' \sigma'} \epsilon_{1\mathbf{k}'} \hat{n}_{1\mathbf{k}' \sigma'} 
 |\Phi_0\rangle
+
 \epsilon_{2\mathbf{k}} \hat{n}_{2\mathbf{k} \sigma} 
 \hat{a}_{2\mathbf{k} \sigma}^\dagger
 |\Phi_0\rangle
\nonumber\\
 &=&
 \left(\sum_{\mathbf{k}'} 2\epsilon_{1\mathbf{k}'} + \epsilon_{2\mathbf{k}}\right)
 |\Phi^e_{2\mathbf{k}\sigma}\rangle,
\end{eqnarray}
where $\hat{H}_\text{t}$ is the first term of Eq. (\ref{Eq:Hnonhyb}).
The second part is calculated as 
\begin{eqnarray}
\hat{H}_\text{bi} 
 |\Phi^e_{2\mathbf{k}\sigma}\rangle 
&=&
 \hat{a}_{2\mathbf{k}\sigma}^\dagger
 \hat{H}_\text{bi} |\Phi_0\rangle
+ [\hat{H}_\text{bi}, \hat{a}_{2\mathbf{k}\sigma}^\dagger]|\Phi_0\rangle,
\end{eqnarray}
where $\hat{H}_\text{bi}$ is the second term of Eq. (\ref{Eq:Hnonhyb}).
The first term is 
\begin{eqnarray}
 \hat{a}_{2\mathbf{k}\sigma}^\dagger
 \sum_{\mathbf{m}} 
  \sum_{\lambda} U_\parallel \hat{n}_{\lambda \mathbf{m} \uparrow} \hat{n}_{\lambda \mathbf{m} \downarrow}
 |\Phi_0\rangle
&=&
 \hat{a}_{2\mathbf{k}\sigma}^\dagger
 N U_{\parallel} |\Phi_0\rangle
\nonumber\\
&=&
 N U_{\parallel} |\Phi_{2\mathbf{k}\sigma}^e\rangle.
\end{eqnarray}
The second term is 
\begin{eqnarray}
 [\hat{H}_\text{bi}, \hat{a}_{2\mathbf{k}\sigma}^\dagger] 
 &=&
 \sum_{\mathbf{m}} \frac{e^{i\mathbf{k}\cdot\mathbf{m}}}{\sqrt{N}} 
 [\hat{H}_\text{bi}, \hat{c}_{2\mathbf{m}\sigma}^\dagger] 
\nonumber\\
 &=&
 \sum_{\mathbf{m}} \frac{e^{i\mathbf{k}\cdot\mathbf{m}}}{\sqrt{N}} 
 \left\{
  U_\parallel [\hat{n}_{2 \mathbf{m} \uparrow} \hat{n}_{2 \mathbf{m} \downarrow}, 
  \hat{c}_{2\mathbf{m}\sigma}^\dagger]
  \right.
\nonumber\\
 &+&
  \left.
  (U_\perp - J) 
   \left(
    \delta_{\uparrow \sigma} 
    \hat{n}_{1\mathbf{m} \uparrow} \hat{c}_{2\mathbf{m} \uparrow}^\dagger
  + 
    \delta_{\downarrow \sigma}
    \hat{n}_{1\mathbf{m} \downarrow} \hat{c}_{2\mathbf{m} \downarrow}^\dagger
   \right)
  \right.
\nonumber\\
 &+&
  U_\perp 
   \left(
    \delta_{\downarrow \sigma}
    \hat{n}_{1\mathbf{m} \uparrow} \hat{c}_{2\mathbf{m} \downarrow}^\dagger
  + 
    \delta_{\uparrow \sigma} 
    \hat{n}_{1\mathbf{m} \downarrow} \hat{c}_{2\mathbf{m} \uparrow}^\dagger
    \right)
\nonumber\\
 &+&
  \left.
   J
   \left(
    \hat{c}_{1\mathbf{m}\uparrow}^\dagger
    \hat{c}_{1\mathbf{m}\downarrow}^\dagger
    \left(
    \delta_{\sigma\downarrow}
    \hat{c}_{2\mathbf{m}\uparrow}
    - 
    \delta_{\sigma\uparrow}
    \hat{c}_{2\mathbf{m}\downarrow}
    \right)
  \right.
  \right.
\nonumber\\
&-&
  \left.
  \left.
    \delta_{\sigma\uparrow}
    \hat{c}_{2\mathbf{m}\downarrow}^\dagger
    \hat{c}_{1 \mathbf{m}\uparrow}^\dagger
    \hat{c}_{1 \mathbf{m}\downarrow}
-
    \delta_{\sigma\downarrow}
    \hat{c}_{2 \mathbf{m}\uparrow}^\dagger
    \hat{c}_{1\mathbf{m}\downarrow}^\dagger
    \hat{c}_{1\mathbf{m}\uparrow}
   \right)
 \right],
\nonumber\\
\end{eqnarray}
and thus, 
\begin{eqnarray}
 [\hat{H}_\text{bi}, \hat{a}_{2\mathbf{k}\sigma}^\dagger] |\Phi_0\rangle 
 &=&
 \sum_{\mathbf{m}} \frac{e^{i\mathbf{k}\cdot\mathbf{m}}}{\sqrt{N}} 
  (2U_\perp - J)\hat{c}_{2\mathbf{m}\sigma}|\Phi_0\rangle 
\nonumber\\
 &=&
  (2U_\perp - J) |\Phi_{2\mathbf{k}\sigma}^e\rangle,
\end{eqnarray}
where $\hat{H}_\text{bi}$ is the second term of Eq. (\ref{Eq:Hnonhyb}).
Therefore, Eq. (\ref{Eq:Phin+1}) is an eigenstate of the Hamiltonian (\ref{Eq:Hnonhyb}) and the eigen energy $E^{e}_{2\mathbf{k}}$ is 
\begin{eqnarray}
 E^e_{2\mathbf{k}} &=& E_0 + \epsilon_{2\mathbf{k}} + 2U_{\perp} - J.
\label{Eq:Ee}
\end{eqnarray}

Similar situation arises when one hole is added. 
The state is written as 
\begin{eqnarray}
 |\Phi^h_{1\mathbf{k}\sigma}\rangle &=& \hat{a}_{1\mathbf{k}\sigma}|\Phi_0\rangle
\label{Eq:Phin-1}
\end{eqnarray}
and the eigenvalue of the Hamiltonian (\ref{Eq:Hnonhyb}) is 
\begin{eqnarray}
 E^h_{1\mathbf{k}} &=& E_0 - \epsilon_{1\mathbf{k}} - U_{\parallel}.
\label{Eq:Eh}
\end{eqnarray}

From the energies with one electron (\ref{Eq:Ee}) and one hole (\ref{Eq:Eh}),
the quasi-particle band gap $\Delta E$ is obtained as:
\begin{eqnarray}
 \Delta E &=& \epsilon_{2\mathbf{k}} - \epsilon_{1\mathbf{k}'} + U_{\perp} - 3J.
\end{eqnarray}
Here, $\mathbf{k}$ and $\mathbf{k}'$ are the $\mathbf{k}$-points where the 
energies of the empty and the filled bands are the minimum and maximum, respectively.

Generalization of the above discussion to the cases with three or more orbitals is straight forward. 
In the case of three orbitals, we assume that orbitals $\lambda = 1, 2$ are doubly occupied and orbital $\lambda = 3$ is empty.
The form of the eigenstates are the same as before. 
The eigenstate are given in the main text.
The energies are 
\begin{eqnarray}
 E_0 &=& \sum_{\mathbf{k}} 2(\epsilon_{1\mathbf{k}} + \epsilon_{2\mathbf{k}})
         + N(2U_{\parallel} + 4U_{\perp} - 2J).
\\
 E_{3\mathbf{k}}^e &=& E_0 + \epsilon_{3\mathbf{k}} + (4U_{\perp} -2J),
\\
 E_{\lambda\mathbf{k}}^h &=& E_0 - \epsilon_{\lambda\mathbf{k}} -
                             (U_{\parallel} + 2U_{\perp} -J).
\end{eqnarray}
Therefore, the energy gap is 
\begin{eqnarray}
 \Delta E &=& \epsilon_{3\mathbf{k}} - \epsilon_{\lambda \mathbf{k}'} + U_\perp - 3J.
\end{eqnarray}
Here, $\mathbf{k}$ and $\lambda\mathbf{k}'$ are taken so that the
energies of the empty and the filled bands become the minimum and maximum, respectively.
These formula hold for the systems with hybridization within the occupied (or empty) bands with slight change. 


%

\end{document}